\documentclass[%
reprint,
superscriptaddress,
amsmath,amssymb,
aps,
pra,
]{revtex4-2}

\usepackage{graphicx}% Include figure files
\usepackage{dcolumn}% Align table columns on a decimal point
\usepackage{float}    % for H option

\usepackage{bm}% bold math
\usepackage{amsthm}
\usepackage{amssymb}
\usepackage{float} 
\usepackage{amsmath} 
\usepackage{physics} 
\usepackage{empheq}
\usepackage[colorlinks=true, linkcolor=blue, citecolor=blue, urlcolor=blue]{hyperref}

\usepackage{xcolor}
\usepackage[normalem]{ulem}
\usepackage{color,soul}
\usepackage{soul}

\begin{document}

\preprint{APS/123-QED}

\title{Spin-free potential-energy curves, spectroscopic constants, and permanent dipole-moment functions for all 18 CO states correlating with the first atomic asymptote}% Force line breaks with \\
%++
\author{Patryk Jasik}\email{patryk.jasik@pg.edu.pl}
\affiliation{Faculty of Applied Physics and Mathematics, Gdańsk University of Technology, Narutowicza 11/12 Street, 80-233 Gdańsk, Poland}
\affiliation{BioTechMed Center, Gdańsk University of Technology, Narutowicza 11/12 Street, 80-233 Gdańsk, Poland}
\affiliation{Centre for Space Biomedicine and Radiological Protection, Medical University of Gdańsk, D\c{e}binki 7 Street, 80-211 Gdańsk, Poland}
\author{Benjamin A. Ikuesan}
\affiliation{Faculty of Applied Physics and Mathematics, Gdańsk University of Technology, Narutowicza 11/12 Street, 80-233 Gdańsk, Poland}
\affiliation{Doctoral School at Gdańsk University of Technology, Gdańsk University of Technology, Narutowicza 11/12 Street, 80-233 Gdańsk, Poland}
\author{Niyazi Bulut}
\affiliation{Faculty of Applied Physics and Mathematics, Gdańsk University of Technology, Narutowicza 11/12 Street, 80-233 Gdańsk, Poland}
\affiliation{Physics Department, Faculty of Science, Firat University, 23119 Elazig, Türkiye}
\author{Stanisław Ryzner}
\affiliation{Materials Spectroscopy Laboratory, Institute of Physics, University of Rzeszów, 1 Pigonia Street, 35-310 Rzeszów, Poland}
\affiliation{Doctoral School of the University of Rzeszów, University of Rzeszów, 16C Rejtana Street, 35-959 Rzeszów, Poland}
\author{Rafał Hakalla}
\affiliation{Materials Spectroscopy Laboratory, Institute of Physics, University of Rzeszów, 1 Pigonia Street, 35-310 Rzeszów, Poland}
\author{Józef E. Sienkiewicz}
\affiliation{Faculty of Applied Physics and Mathematics, Gdańsk University of Technology, Narutowicza 11/12 Street, 80-233 Gdańsk, Poland}
\affiliation{Advanced Materials Center, Gdańsk University of Technology, Narutowicza 11/12 Street, 80-233 Gdańsk, Poland}

\date{\today}% It is always \today, today,

\begin{abstract}

%\begin{description}
%You may use the %\texttt{description} environment to structure your abstract;
%use the optional argument of the %\verb+\item+ command to give the category of each item. 
%\end{description}

% 
We present a systematic \textit{ab initio} characterization of the eighteen spin-free adiabatic electronic states of $^{12}$C$^{16}$O that correlate with the first dissociation asymptote $\mathrm{C}(^{3}P)+\mathrm{O}(^{3}P)$. Complementing the set of potential-energy curves recently reported by Bop and Gacesa, we calculate the full singlet, triplet, and quintet manifold at the state-averaged CASSCF/icMRCI(+Q) level using the aug-cc-pV5Z basis set. Basis-set convergence is examined in detail for the ground electronic state through calculations with the aug-cc-pV$n$Z ($n=T, Q, 5, 6$) hierarchy and complete-basis-set limit extrapolation. The resulting potential-energy curves reproduce the established topology of the low-lying CO states, including the double-minimum structure of the 2$^{1}\Sigma^{+}$ state and pronounced adiabatic mixing in several excited states. Rovibrational levels and spectroscopic constants are determined for the bound states and characterized local minima, and are compared with available experimental measurements and previous theoretical results. Good agreement is obtained for the ground state and the experimentally well-characterized low-lying singlet and triplet states. At the same time, predictions are provided for less thoroughly studied quintet and higher-lying states. For the ground state 1$^{1}\Sigma^{+}$, we obtain R$_e$ = 1.131~$\mathrm{\AA}$ and $\omega_e$ = 2161.04~$\mathrm{cm^{-1}}$, both within 0.5\% of the corresponding experimental values, together with a bond dissociation energy D$_0$ of 89473~$\pm$~ 141~$\mathrm{cm^{-1}}$. Among the quintet states, 1$^{5}\Pi$ is the most strongly bound, with D$_e$ = 6401 and $\omega_e$ = 982~$\mathrm{cm^{-1}}$. By contrast, 1$^{5}\Sigma^{+}$ supports only a shallow long-range well with D$_e$ = 849~$\mathrm{cm^{-1}}$, whereas 2$^{5}\Sigma^{+}$, 2$^{5}\Pi$, and 1$^{5}\Delta$ are predominantly repulsive. In addition, permanent dipole-moment functions are reported for all eighteen states over the investigated range of internuclear separations. Their variations with bond length reveal changes in electronic character and charge redistribution, particularly near regions of strong state interaction. The comprehensive analysis of the uncertainties of the reported results gives insight into the reliability and consistency of our calculations.  The present results provide an independent, internally consistent spectroscopic and electric-property characterization of 18 electronic states of the CO molecule and complement existing potential-energy data for applications in rovibrational spectroscopy and future nonadiabatic, spin--orbit-coupled, and collision-dynamics studies.

\end{abstract}

\maketitle

%\tableofcontents

\section{Introduction}

Carbon monoxide (CO) is, after molecular hydrogen, one of the most abundant molecular species in space and one of the most extensively observed tracers of the physical and chemical conditions prevailing in interstellar and circumstellar environments \cite{Visser2009, Bolatto2013}. It is also an important constituent of planetary atmospheres, cometary comae, and combustion systems, in which its spectroscopic signatures provide quantitative information on temperature, density, chemical composition, and dynamical processes. In the interstellar medium, CO constitutes a major reservoir of carbon and participates in the formation of more complex molecular species through both gas-phase reaction pathways and chemical processes occurring on the surfaces of interstellar dust grains. The central role of CO in interstellar clouds and circumstellar disks was reviewed by Visser et al. \cite{Visser2009}, who demonstrated that its gas-phase abundance is strongly controlled by ultraviolet photodissociation and molecular self-shielding.

A quantitatively reliable description of CO under the broad range of physical conditions encountered in these environments requires accurate potential-energy curves, spectroscopic parameters, and permanent dipole-moment functions for its electronically excited states \cite{Ushakov2020}. The PECs and spectroscopic parameters characterize the energy-level structure, while the permanent dipole-moment functions provide the dipole information needed to calculate electric-dipole rotational and rovibrational transition intensities within a given electronic state~\cite{Li2015CO, Spirko2024}. Recent state-resolved studies of vibrational energy transfer in CO-O$_2$ collisions have further demonstrated the importance of accurate potential-energy representations for distinguishing competing vibration-to-vibration and vibration-to-translation/rotation mechanisms and for constructing reliable kinetic models over broad temperature ranges \cite{Coletti2026}. Moreover, the relatively compact electronic structure of CO, together with the extensive body of available experimental data, renders this molecule a particularly stringent benchmark for assessing the accuracy, convergence, and internal consistency of high-level electronic-structure methods \cite{Gong2023}.

Despite intensive experimental and theoretical investigations \cite{Farrenq1991,Tobias1960,Krupenie1965,Jimeno1998,Vazquez2009,Kumar2010,Liu2011,Lu2013,Powell2016,Xing2018,Abdulah2019,Coe2019,Araujo2023}, the excited-state electronic structure of CO remains challenging to characterize. In the vacuum-ultraviolet region, several electronic bands, particularly those involving states of ${}^{1}\Pi$ symmetry, display strong perturbations caused by interactions among densely spaced valence and Rydberg states. Considerable progress has been made in identifying and deperturbing these structures through high-resolution spectroscopy and complementary theoretical analysis \cite{FIELD1972, Niu2013, Hakalla2019, Malicka2020, Malicka2021, Ryzner2022, Malicka2024}. Nevertheless, avoided crossings, rapid changes in electronic character, and the coexistence of bound, weakly bound, and locally trapped states continue to complicate the construction of a uniform electronic-structure description over an extended range of internuclear separations.

In the spin-free representation, the interaction of $\mathrm{C}(^{3}P)$ and $\mathrm{O}(^{3}P)$ gives rise to 18 molecular electronic terms: two states of each ${}^{1,3,5}\Sigma^{+}$ and ${}^{1,3,5}\Pi$ symmetry, and one state of each ${}^{1,3,5}\Sigma^{-}$ and ${}^{1,3,5}\Delta$ symmetry. Earlier theoretical studies generally considered selected subsets of these electronic states, with particular emphasis on the experimentally accessible low-lying singlet and triplet states. Bauschlicher, Langhoff, and Partridge investigated the six quintet states \cite{Bauschlicher1993}, while more recent calculations provided extensive data for the singlet and triplet manifolds \cite{Khalil2025}. Recently, Bop and Gacesa \cite{Bop2025} reported a set of all 18 PECs calculated at the MRCISD+Q level with an augmented correlation-consistent quintuple-zeta basis set. Their potentials were extended into the long-range region and employed in calculations of elastic $\mathrm{C}({}^{3}P)+\mathrm{O}({}^{3}P)$ scattering cross sections. That work established an important complete reference for the topology and asymptotic behaviour of the 18-state manifold. The present work complements this reference by providing permanent dipole-moment functions for all eighteen electronic states, together with electronic and rovibrational spectroscopic constants for the bound states and characterised local minima. We also assess computational uncertainties in the PECs and derived spectroscopic constants, including basis-set incompleteness and uncertainties from the rovibrational fitting procedure. For the ground state, we employ two complementary CBS extrapolation approaches (inverse-power fitting and stochastic random-walk analysis) and report both the CBS-limit energy at the equilibrium geometry and the CBS-extrapolated potential-energy curve, with approximately 95\% confidence intervals obtained from the stochastic analysis and evaluated pointwise along the curve. Very recently, Zhang et al. \cite{Zhang2026} reported rovibronic spectra for eight singlet and triplet electronic states of CO, $1^{1}\Sigma^{+}, 1^{1}\Pi, 1^{1}\Sigma^{-}, 1^{1}\Delta, 1^{3}\Pi, 1^{3}\Delta, 1^{3}\Sigma^{+}, 1^{3}\Sigma^{-}$, using hybrid RKR/\textit{ab initio} potential-energy curves that include spin--orbit and electronic angular-momentum coupling matrix elements. That study focused on vacuum-ultraviolet absorption and emission spectra at high temperatures, $3000\text{--}8000~\mathrm{K}$, with particular emphasis on applications to stellar-atmosphere modeling and interstellar radiative transfer. Although it provided detailed spectroscopic transition data for the singlet--triplet states, it did not consider $2^{1,3}\Sigma^+$, $2^{1,3}\Pi$, and the six quintet states or report permanent dipole-moment functions for the complete electronic-state manifold.

The present work, therefore, pursues a complementary objective: a uniform, spin-free \textit{ab initio} characterization of all 18 electronic states correlating with the first dissociation limit, together with their
rovibrational constants and permanent dipole-moment functions. The resulting electronic-structure dataset provides a foundation for future nonadiabatic and spin--orbit-coupled dynamical simulations.

The availability of complete PECs, however, does not remove the need for a uniform characterization of the associated spectroscopic and electrical properties. In particular, rovibrational constants and permanent dipole-moment functions have not been documented with the same level of consistency for the entire 18-state manifold. Permanent dipole-moment functions provide complementary information on the redistribution of electronic charge with internuclear separation and are sensitive to changes in the composition of adiabatic electronic wavefunctions. Together with the PECs, they are required for calculating intrastate rovibrational dipole matrix elements and can assist in identifying regions of strong electronic mixing. Independent calculations using a common orbital and correlation framework are also useful for assessing the robustness of previously reported potential features and for identifying states and geometries at which the theoretical description is particularly sensitive to basis-set, active-space, or state-averaging choices.

In the present work, we therefore perform a systematic calculation of the complete set of 18 spin-free electronic states correlating with the $\mathrm{C}(^{3}P)+\mathrm{O}(^{3}P)$ dissociation limit. The PECs are obtained using state-averaged complete-active-space self-consistent-field wavefunctions followed by internally contracted multireference configuration interaction with the Davidson correction, CASSCF/icMRCI(+Q), in conjunction with the aug-cc-pV5Z basis set. Basis-set convergence is examined in detail for the $1{}^{1}\Sigma^{+}$ ground state using the aug-cc-pV$n$Z $(n=D,T,Q,5,6)$ hierarchy and complete-basis-set extrapolation. The resulting curves are compared with available experimental information, previous theoretical calculations, and the recent PECs \cite{Bop2025, Zhang2026}.

Rovibrational levels are calculated for the bound states and characterized local minima, from which spectroscopic constants are extracted and compared with experimental and theoretical reference values. In addition, molecule-fixed permanent dipole-moment functions are reported for all eighteen states over the investigated range of internuclear separations. Particular attention is given to their behavior in regions of strong adiabatic mixing and to comparisons with available dipole data for the $1{}^{1}\Sigma^{+}$ and $1{}^{3}\Pi$ states. The purpose of the present study is thus not to claim the first construction of the eighteen PECs, but to provide an independent and internally consistent spectroscopic and electric-property characterization that complements the existing long-range potential and scattering results.

The paper is organized as follows. Section~II describes the electronic-structure calculations, active-space and state-selection procedures, basis-set convergence analysis, and rovibrational calculations. Section~III presents and discusses the potential-energy curves, spectroscopic constants, and permanent dipole-moment functions. Section~IV examines the principal sources of uncertainty in the calculated energies and derived molecular constants. Finally, Section~V summarizes the main conclusions and outlines possible extensions involving transition properties, nonadiabatic interactions, and spin--orbit coupling. The numerical data accompanying the present work are described in the Data Availability statement.

\section{Computational method}

\subsection{Methods of electronic-structure calculations}

The electronic structure of $^{12}$C$^{16}$O was calculated using a complete active space self-consistent field (CASSCF) reference, followed by the multireference configuration interaction method with the Davidson correction (icMRCI(+Q)), as implemented in \textsc{MOLPRO 2023} \cite{Werner1985, Werner1988MRCI, Zhang2021, molpro2023}. 

State-averaged CASSCF (SA-CASSCF) was employed to obtain a common set of orbitals for the targeted states, which in the present work included singlet, triplet, and quintet states; equal weights were assigned to avoid bias toward any single state. Dynamical electron correlation was then recovered at the icMRCI level. The energies were supplemented with the relaxed-reference Davidson correction implemented in MOLPRO. This a posteriori correction approximately accounts for higher-excitation contributions omitted from the icMRCI expansion and reduces the associated size-extensivity error. Earlier applications of the SA-CASSCF/icMRCI(+Q) methodology described above are reported in our previous studies \cite{Jasik2021, Baczek18102022, Jasik2023}. This method provides a consistent, high-accuracy description over a wide range of internuclear separations, suitable for deriving spectroscopic constants and analyzing state interactions.

\subsection{Active space}\label{active_space_section}

To describe the complete set of 18 spin-free electronic states which correlates with C($^3$P) + O($^3$P) asymptote and comprises (1,2)$^1\Sigma^+$, (1,2)$^3\Sigma^+$, (1,2)$^5\Sigma^+$, 1$^1\Sigma^-$, 1$^3\Sigma^-$, 1$^5\Sigma^-$, (1,2)$^1\Pi$, (1,2)$^3\Pi$, (1,2)$^5\Pi$, 1$^1\Delta$, $1^3\Delta$, 1$^5\Delta$, we employed an SA-CASSCF reference with ten active electrons distributed among eleven active spatial orbitals, denoted CAS(10,11). In $C_{2v}$ symmetry, these active orbitals comprise five $a_1$, three $b_1$, and three $b_2$ orbitals. Four additional electrons occupy two inactive, doubly occupied $a_1$ orbitals associated with the C(1s) and O(1s) shells. In MOLPRO, this orbital partition corresponds to OCC = (7,3,3,0) with the default CLOSED = (2,0,0,0), giving eleven active and two inactive orbitals. The two inactive orbitals were optimised together with the active orbitals during SA-CASSCF. The active-space selection was guided by the need for a common description of all eighteen electronic states correlating with C($^3$P) + O($^3$P) asymptote over
the investigated range of internuclear separations, including regions where their electronic character changes. We therefore sought flexibility in both the $\sigma$ and $\pi$ orbital sectors, retaining equal numbers of $b_1$ and $b_2$ orbitals to represent both $\pi$ components. The choice was assessed across the complete target manifold rather than for selected low-lying states alone.

In the subsequent icMRCI calculations, the same CAS(10,11) reference space for the CO molecule was used, together with the default CORE = (2,0,0,0) setting. Thus, the ten valence electrons were correlated, whereas the four electrons occupying the two core orbitals associated with the C(1s) and O(1s) shells were excluded from the correlation treatment. The molecular orbitals obtained from SA-CASSCF were used without further orbital optimisation during icMRCI.

We also tested CAS(10,10) and CAS(10,12) reference spaces, with active-orbital distributions of (6,2,2,0) and (6,3,3,0), respectively, over the ($a_1$,$b_1$,$b_2$,$a_2$) irreducible representations of $C_{2v}$. In MOLPRO, these correspond to OCC = (8,2,2,0) and OCC = (8,3,3,0), respectively, with the default CLOSED = (2,0,0,0) in both cases. Relative to CAS(10,10), the adopted CAS(10,11) reference contains one fewer $a_1$ orbital and one additional orbital in each of the $b_1$ and $b_2$ representations. CAS(10,12), in turn, retains the same $\pi$ orbital space as CAS(10,11) and adds one $a_1$ orbital. The CAS(10,10) reference yielded satisfactory results for the $\Sigma$ states and some low-lying $\Pi$ states, but convergence difficulties arose for higher-lying quintet $\Pi$ states and some singlet $\Pi$ states at certain internuclear separations. Calculations with the larger CAS(10,12) reference exhibited discontinuities in some of the resulting PECs. The difficulties encountered with CAS(10,10) may partly reflect the reduced flexibility of its smaller $\pi$ orbital space. For CAS(10,12), possible causes of the discontinuities include convergence to competing orbital-optimisation solutions or changes in the electronic state followed between neighbouring geometries. Among the spaces examined in the initial PEC scans, CAS(10,11) provided consistent convergence and smooth PECs for all eighteen considered electronic states over the investigated internuclear range and was therefore adopted. This selection was based on the numerical performance obtained with the present computational protocol. These tests do not establish convergence of the PECs or PDM functions with respect to further enlargement of the active $\pi$ orbital space. In our calculations, at both the SA-CASSCF and icMRCI levels, we also included one or two additional higher-lying electronic states beyond those correlated to the C($^3$P) + O($^3$P) asymptote. These additional excited electronic states facilitated the identification of the considered states and the examination of possible interactions with nearby higher-lying states of the same spin and linear-molecule symmetry.

For each $\Pi$ and $\Delta$ electronic states, both $C_{2v}$ symmetry components were included in the calculations to represent its twofold orbital degeneracy. In the $C_{2v}$ representation, $\Sigma^+$ states and one component of each $\Delta$ state belong to $A_1$, whereas $\Sigma^-$ states and the other $\Delta$ component belong to $A_2$. The electronic-state assignments were therefore checked at every calculated internuclear separation using the SA-CASSCF expectation values of $\hat{L}_z^2$. For states with well-defined $\Lambda$, these expectation values are $\Lambda^2\hbar^2$, corresponding to $0$, $1$, and $4$ in atomic units for $\Sigma$, $\Pi$, and $\Delta$ states, respectively. Additional qualitative spot checks were performed at selected internuclear separations by inspecting the state-averaged CASSCF natural-orbital occupation numbers and the leading CI coefficients for individual electronic states. These checks were used to identify unexpected changes in electronic character, taking into account possible configuration mixing near avoided crossings. These checks provided a qualitative assessment of the consistency of the calculated states. The PECs were calculated on a nonuniform grid of 38 internuclear separations covering $R = 0.80$--$4.00$~\AA, with a spacing of $\Delta R = 0.05$~\AA{} between $0.80$ and $1.30$~\AA{} and $\Delta R = 0.10$~\AA{} between $1.30$ and $4.00$~\AA.

\subsection{Basis sets and CBS limit}

The PECs for all eighteen states were calculated using the aug-cc-pV5Z basis set augmented with diffuse orbitals \cite{Dunning1989, Kendall1992}. The systematic basis-set convergence analysis and CBS extrapolations were restricted to the ground electronic state, with aug-cc-pVTZ, aug-cc-pVQZ, aug-cc-pV5Z, and aug-cc-pV6Z results used for the extrapolations. No corresponding basis-set sequence was calculated for the excited-state PECs. At each basis level, we used the same SA-CASSCF reference space, CAS(10,11) for the CO molecule, and the same frozen-core icMRCI(+Q) treatment described above, ensuring a consistent electronic-structure description across basis sizes. Therefore, CBS estimates were obtained with an inverse-power extrapolation model adjusted to the set of calculated total energies,
\begin{equation}
E_{\mathrm{tot}}(n)=E_\infty^{\mathrm{CBS}}+\frac{C}{n^{\alpha}},
\end{equation} \\
where $n$ is the cardinal number ($\mathrm{T}=3$, $\mathrm{Q}=4$, $5$, $6$), and $C$ and $\alpha$ are the fitting parameters. In practice, the extrapolation was performed using the $\{\mathrm T,{Q},5\}$ and $\{\mathrm T,{Q},5,6\}$ basis-set sequences; the procedure follows standard protocols that reduce basis-set incompleteness error and yield more reliable \textit{ab initio} energies for CO at the CBS limit \cite{Peterson1995, Helgaker1997}.

\subsection{Calculations of rovibrational spectroscopic constants}

We used nonrepulsive PECs of the electronic states of the $^{12}$C$^{16}$O molecule to calculate the rovibrational wavefunctions and corresponding energy levels for varying numbers of vibrational levels $v$,  depending on the depth of each PEC, and $J=0-10$. All wavefunctions and energy levels were obtained using the sPYtroscopy program \cite{Jasik2025_spytroscopy}, which implements the discrete variable representation method in the Fourier sine basis.

The spectroscopic constants such as $\omega_e$, $\omega_e x_e$, $B_e$, and $\alpha_e$ were extracted from the calculated rovibrational energies by means of least-squares fits performed with the Python \emph{curve\_fit} function from the \emph{scipy.optimize} module \cite{2020SciPy-NMeth}. For each vibrational quantum number $v$, the calculated rotational ladder was represented by
\begin{equation}
E(v,J) = G_v + B_v X_J - D_v X_J^{2} + H_v X_J^{3} + \cdots,
\end{equation}
where $X_J = J(J+1)$, $G_v$ is the rotationless vibrational term value, $B_v$ is the vibrationally dependent rotational constant, and $D_v$ and $H_v$ are the quartic and sextic centrifugal-distortion constants, respectively. In the present analysis, the lowest-order model that reproduced the calculated rovibrational energies to the numerical precision was employed. In most cases, the expression $E(v,J)=G_v + B_v X_J - D_v X_J^{2}$ was sufficient. The sextic term was included only when a systematic dependence of the residuals on $J$ remained after inclusion of the quartic centrifugal-distortion contribution. Fitting the intercept $G_v$, rather than subtracting an energy at a particular $J$, also makes the procedure applicable to electronic states for which $J=0$ is not an allowed rotational level (e.g., $\Pi$ or $\Delta$ states).

The truncated vibrational Dunham expansion subsequently represented the vibrational term values obtained from the rotational fits
\begin{equation}
G_v = G_{\mathrm{ref}} +\omega_e\left(v+\frac{1}{2}\right) 
                       -\omega_ex_e\left(v+\frac{1}{2}\right)^2 
                       +\omega_ey_e\left(v+\frac{1}{2}\right)^3 ,
\end{equation}
where $G_{\mathrm{ref}}$ accounts for the selected electronic-energy reference. The fitted coefficients directly provide the harmonic vibrational constant $\omega_e$, the vibrational anharmonicity constant $\omega_ex_e$, and the higher-order anharmonicity constant $\omega_ey_e$. The inclusion of $G_{\mathrm{ref}}$ makes the fitted spectroscopic constants independent of the arbitrary energy zero used for the underlying electronic potential.

The vibrational dependence of the rotational constants was fitted independently according to
\begin{equation}
B_v = B_e - \alpha_e\left(v+\frac{1}{2}\right) + \gamma_e\left(v+\frac{1}{2}\right)^2,
\end{equation}
where $B_e$ is the equilibrium rotational constant, $\alpha_e$ describes the first-order vibration–rotation interaction, and $\gamma_e$ represents its leading higher-order correction. Accordingly, $B_e$, $-\alpha_e$, and $\gamma_e$ correspond to the constant, linear, and quadratic coefficients, respectively, of the fitted dependence of $B_v$ on $v+\tfrac12$.

Only the low-lying vibrational and rotational levels, for which the truncated Dunham expansions remain applicable, were used in the final fits. The robustness of the extracted constants was examined by systematically varying the maximum vibrational and rotational quantum numbers included in the analysis. A fit was regarded as stable when the addition or removal of the highest included level did not produce changes in the spectroscopic constants exceeding their estimated fitting uncertainties and did not introduce a systematic structure into the residuals. This procedure limits contamination of the equilibrium spectroscopic constants by strongly anharmonic levels near dissociation or by high-$J$ levels that require centrifugal-distortion terms beyond those retained in the fitting model. The higher-order constants $\omega_e y_e$ and $\gamma_e$ are therefore not reported in the present work, because truncating the expansions of $G_v$ and $B_v$ after four and three terms, respectively, may render the highest-order fitted coefficients particularly susceptible to large uncertainties, strong parameter correlations, and numerical instabilities.

For each least-squares fit, \emph{curve\_fit} function returns the optimized parameter vector, $\mathbf{p}_{\mathrm{opt}}$, and the estimated covariance matrix of the fitted parameters, $\mathbf{C}_p$. Here, $\mathbf{p}_{\mathrm{opt}}$ denotes the vector containing the best-fit values of all adjustable model parameters, $p_i$ is the $i^{th}$ fitted parameter, and $\mathbf{C}_p$ is the covariance matrix describing the estimated variances and mutual correlations of the fitted parameters. The formal one-standard-deviation uncertainty of $p_i$ was calculated from the corresponding diagonal element of the covariance matrix as
\begin{equation}
u_{\mathrm{fit}}(p_i) = \sqrt{\left(\mathbf{C}_p\right)_{ii}},
\end{equation}
where $\left(\mathbf{C}_p\right)_{ii}$ is the estimated variance of the parameter $p_i$.

\section{Results and discussion}

This section analyzes the calculated adiabatic PECs and PDM functions for the 18 lowest-lying electronic states of CO that, at large internuclear separation, correlate with the first atomic dissociation asymptote $\mathrm{C}(^3P)+\mathrm{O}(^3P)$. All calculations were done for the most abundant isotopes of carbon ($^{12}$C) and oxygen ($^{16}$O) atoms. The present calculations provide a consistent description of all states in the $\Sigma$ (both $+$ and $-$), $\Pi$, and $\Delta$ symmetries across the investigated range of bond lengths, thereby enabling a uniform comparison of potential-energy curves within a single theoretical framework.

A complete set of PECs for these eighteen states was recently reported by Bop and Gacesa \cite{Bop2025}. Their calculations employed the MRCISD+Q method with an augmented correlation-consistent quintuple-zeta basis set and extended the potentials into the long-range region required for atomic-scattering calculations. The present curves reproduce the principal topological features of that complete dataset and of earlier state-specific calculations \cite{Bauschlicher1993, Khalil2025}. Because the Bop and Gacesa potentials were primarily constructed for long-range collision dynamics, whereas the present work focuses on the bonding and intermediate-distance regions relevant to molecular spectroscopy and dipole properties, the two datasets are complementary. In particular, the present calculations provide the electronic potentials used consistently to determine rovibrational constants and permanent dipole-moment functions for the entire eighteen electronic states.
\begin{figure}
    \centering
    \includegraphics[width=\columnwidth]{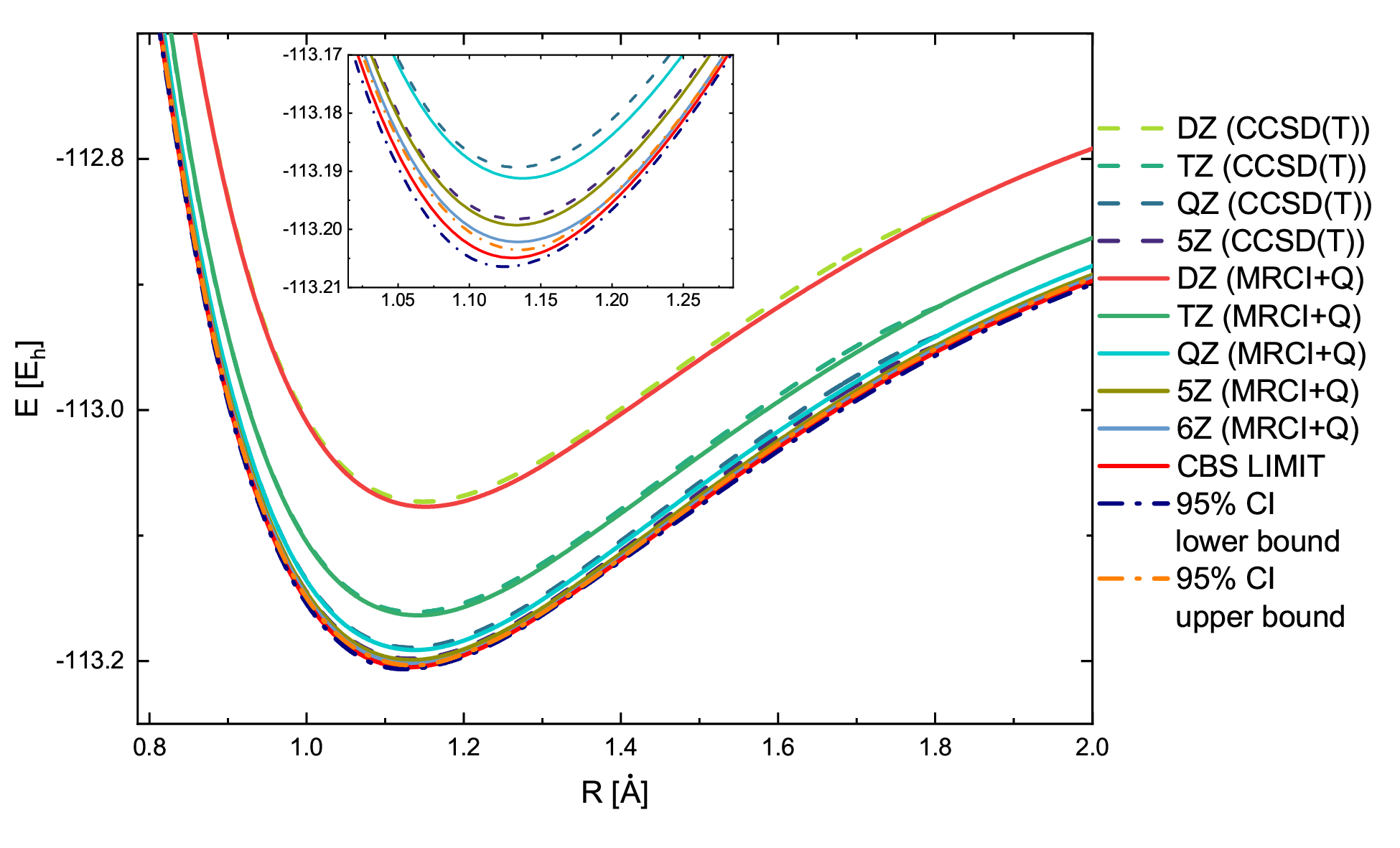}
     \caption{
    Basis-set convergence of the CO ground-state potential-energy curve obtained using CCSD(T) and icMRCI(+Q) calculations with the aug-cc-pV$n$Z basis-set sequence. The CBS extrapolated curve is shown together with the lower and upper confidence bounds. The inset magnifies the equilibrium region, where the aug-cc-pV5Z icMRCI(+Q) PEC lies close to both the aug-cc-pV6Z and CBS-extrapolated curves. Dashed lines represent the CCSD(T) results, solid lines represent the icMRCI(+Q) results and CBS limit, while dash-dotted lines mark the lower and upper bounds of the pointwise 95\% confidence intervals, corresponding to the 2.5th and 97.5th percentiles of the random-walk distributions.
    }
    \label{fig:CCSDT}
\end{figure}

To provide a complementary single-reference benchmark for the ground electronic state, additional CCSD(T) calculations were carried out with the PSI4 program package \cite{Smith2020}, using the aug-cc-pV$n$Z basis set (with particular emphasis on aug-cc-pV5Z). Figure~\ref{fig:CCSDT} shows the systematic basis-set convergence of the CO ground-state potential-energy curve obtained using CCSD(T) and icMRCI(+Q) calculations. The potentials gradually approach the ground state calculated in the complete basis set (CBS) limit as the basis-set size increases, as shown in the inset highlighting the equilibrium region. The close agreement of the aug-cc-pV5Z and aug-cc-pV6Z icMRCI(+Q) curves with the CBS extrapolated curve indicates that the calculated ground state PECs are affected by a very small uncertainty connected with the basis-set incompleteness errors. The 95\% confidence interval provides an estimate of the uncertainty for the ground-state PEC calculated in the CBS limit. The pointwise 95\% confidence bounds for the CBS-extrapolated ground-state PEC shown in Fig. \ref{fig:CCSDT} were obtained using the stochastic random-walk procedure described in Section \ref{sec:uncertainty}. The CBS energies and uncertainty estimates at the equilibrium geometry are summarised in Table \ref{tab:cbs_rw}. The ground-state convergence study provides a benchmark for basis-set effects within the common computational protocol used for all eighteen states. Its relevance to the excited-state PECs is assessed by analogy and supplemented by the comparisons of state-specific spectroscopic constants in Table~\ref{tab:spectro}; a corresponding basis-set sequence was not calculated for the excited states.

\subsection{Potential-Energy Curves}

\begin{figure} % one column only
    \centering
    \includegraphics[width=\columnwidth]{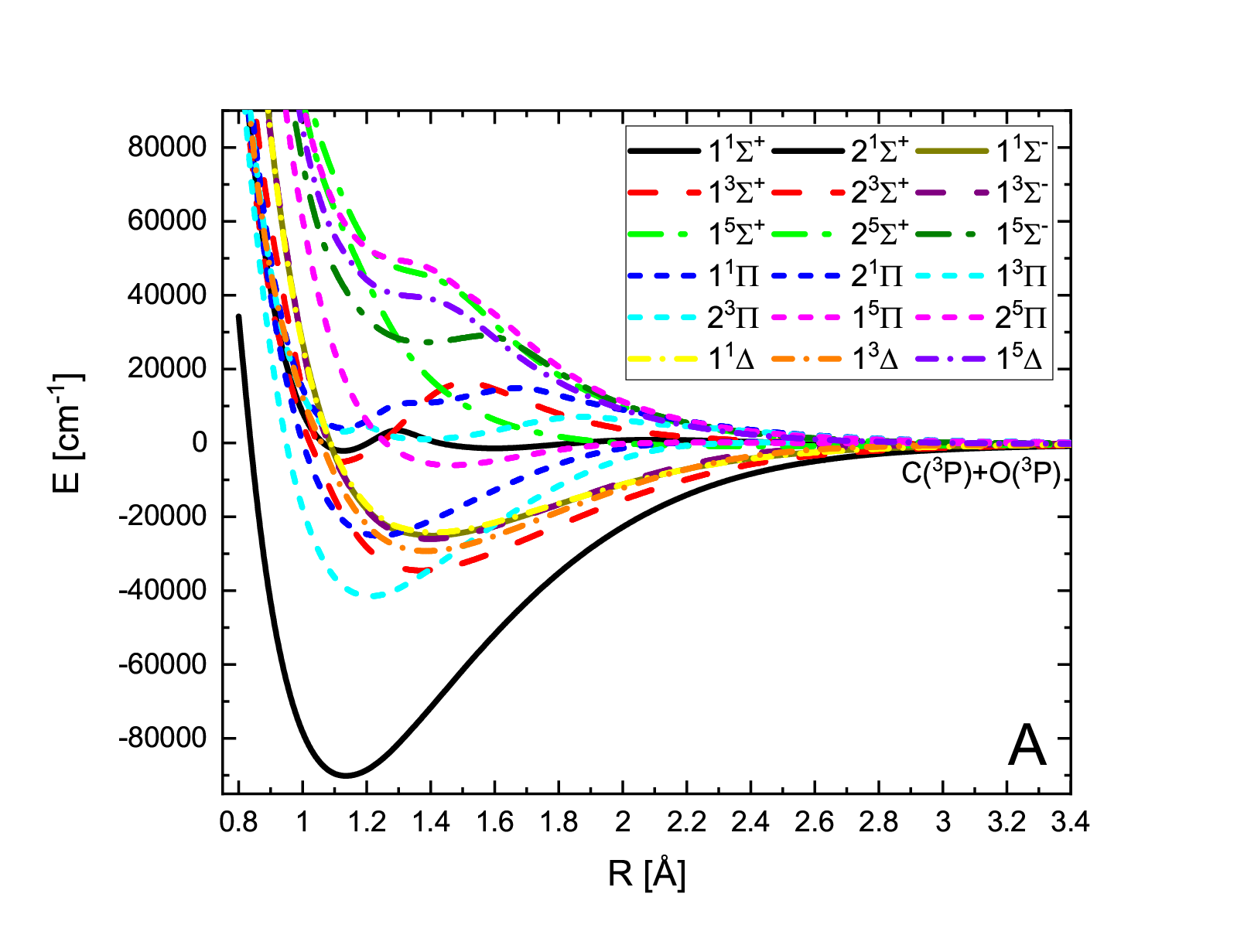}\\
    \vspace{-1.0cm}
    \includegraphics[width=\columnwidth]{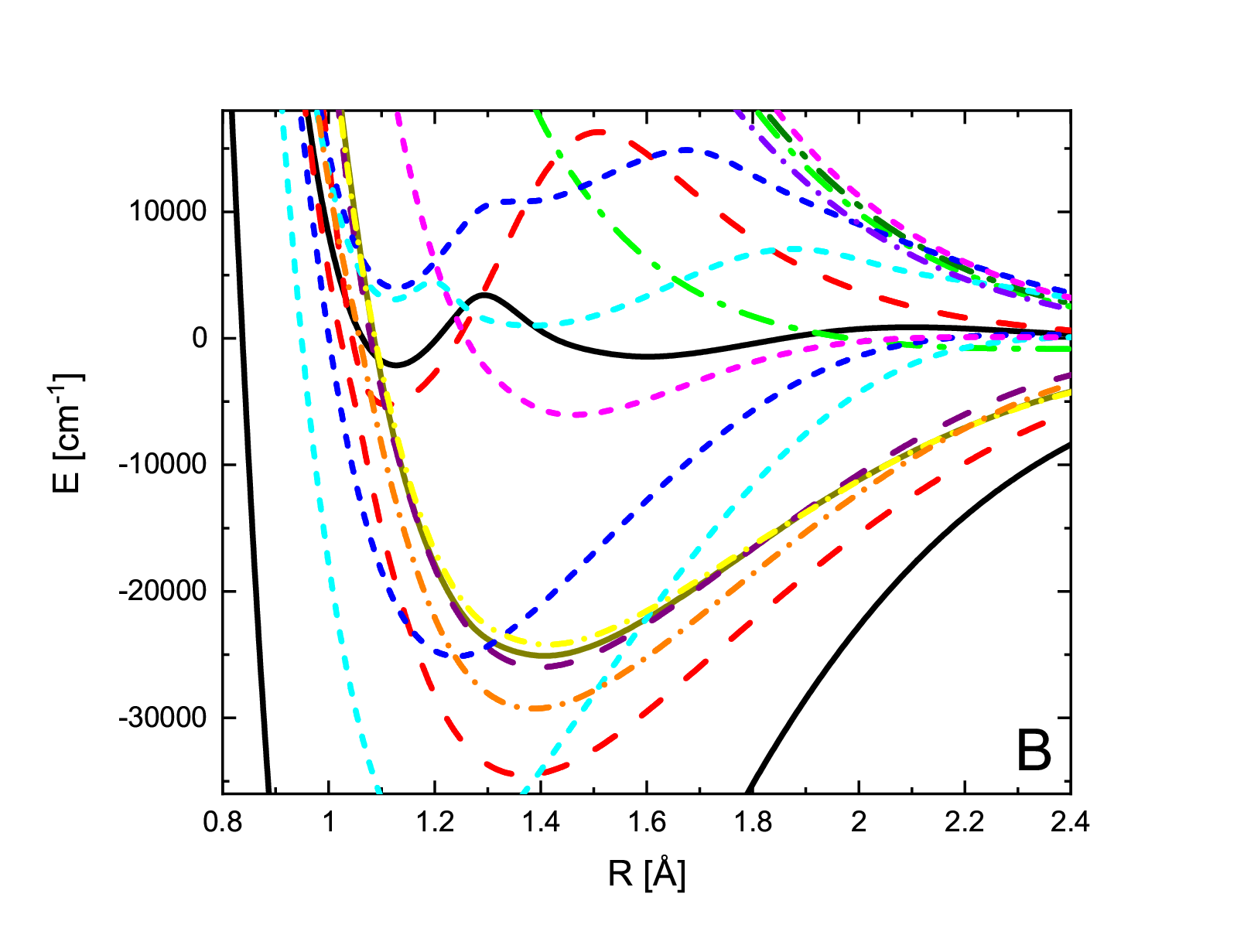}
    \caption{Potential-energy curves of the complete set of 18 electronic states of the CO molecule correlating with the C($^3$P)+O($^3$P) asymptote. Panel A shows the overall arrangement of the potentials, while panel B provides an expanded view over the energy range from -36000 to 18000 cm$^{-1}$, emphasising the relative positions of the curves. The curves are identified as follows. Solid lines: (1,2)$^1\Sigma^+$, black; 1$^1\Sigma^-$, dark yellow. Dashed lines: (1,2)$^3\Sigma^+$, red; 1$^3\Sigma^-$, purple. Dash-dotted lines: (1,2)$^5\Sigma^+$, green; 1$^5\Sigma^-$, olive. Short-dashed lines: (1,2)$^1\Pi$, blue; (1,2)$^3\Pi$, cyan; (1,2)$^5\Pi$, magenta. Short-dash-dotted lines: 1$^1\Delta$, yellow; $1^3\Delta$, orange; 1$^5\Delta$, violet. Within each group labelled (1,2), both states share the indicated colour and line style. The same conventions apply to both panels.}
    \label{fig:PEC}
\end{figure}

Figure \ref{fig:PEC} presents the PECs for the eighteen electronic states specified in Section \ref{active_space_section}.

The $1^{1}\Sigma^{+}$ ground-state curve (solid black line) possesses the deepest potential energy well. It reflects the strong chemical bond in CO. Its equilibrium region is well separated from the remaining curves. Its shape is consistent with the basis-set convergence behaviour shown in Figure \ref{fig:CCSDT}. The low-lying excited singlet states $1^{1}\Sigma^{-}$ (solid dark yellow line), $1^{1}\Pi$ (short-dashed blue line), and $1^{1}\Delta$ (short-dash-dotted yellow line) also exhibit well-defined minima and correspond to the experimentally known results \cite{HuberHerzberg1979}.

The $2^{1}\Sigma^{+}$ curve (solid black line) displays a pronounced double-minimum structure. This feature has been discussed in previous experimental and theoretical studies \cite{HuberHerzberg1979, Shang2026, Eidelsberg1987, wolk1983, Khalil2025} and is also reproduced by the PECs set of Bop and Gacesa \cite{Bop2025}. In the present work, the two minima are treated separately when their local spectroscopic properties are analysed; the resulting constants should therefore be understood as local-well parameters of the adiabatic potential.

Within the triplets, the $1^{3}\Pi$ state (short-dashed cyan line) forms the lowest excited-state minimum. The $1^{3}\Sigma^{+}$ (dashed red line), $1^{3}\Delta$ (short-dash-dotted orange line), and $1^{3}\Sigma^{-}$ (dashed purple line) curves are also strongly bound over the spectroscopically relevant region. The higher $2^{3}\Sigma^{+}$ (dashed red line) and $2^{3}\Pi$ (short-dashed cyan line) curves display shallower and more strongly perturbed minima. Their rapidly varying curvature reflects changes in adiabatic electronic character. This makes the derived local spectroscopic parameters more sensitive to the selected fitting interval than those of the lower triplet states. The $1^{3}\Sigma^{+}$ and $2^{3}\Sigma^{+}$ PECs exhibit an avoided crossing in the short-range repulsive region near $R=1.1$~\AA. A separate interaction of $2^{3}\Sigma^{+}$ with
a higher-lying $^{3}\Sigma^{+}$ state gives rise to the pronounced local maximum in its PEC near $R=1.6$~\AA. The associated variations in the PDM functions are discussed in Section~\ref{subsection_PDM}.

For the $2^{3}\Pi$ state, the broad, shallow potential well, characterised by $\omega_e$ = 711~$\mathrm{cm}^{-1}$ and $\omega_{e}x_{e}$ = -65~$\mathrm{cm}^{-1}$, together with the relatively large uncertainties (51 and 15 $\mathrm{cm}^{-1}$, respectively) in the derived spectroscopic constants, indicates the presence of a nearby avoided crossing with the $3^{3}\Pi$ potential-energy curve. The electron-correlation treatment and the state-averaging weights are likely responsible for the discrepancy between the calculated and experimental $T_e$ values for the $2^{3}\Pi$ state listed in Table~\ref{tab:spectro}.

The quintet curves exhibit a combination of weakly attractive, locally trapped, and predominantly repulsive behaviour. The $1^{5}\Sigma^{+}$ state (dash-dotted green line) possesses a shallow minimum at an extended bond length, whereas the $1^{5}\Pi$ curve (short-dashed magenta line) displays a more compact local well. By contrast, the $2^{5}\Sigma^{+}$ (dash-dotted green line), $2^{5}\Pi$ (short-dashed magenta line), and $1^{5}\Delta$ (short-dash-dotted violet line) states are predominantly repulsive over the investigated interval. The $1^{5}\Sigma^{-}$ PEC (dash-dotted olive line) contains a local minimum above the lowest dissociation limit and therefore does not represent a conventionally bound electronic state. The predominantly repulsive or weakly bound character of most quintet states arises from the high-spin coupling of the $\mathrm{C}(^{3}P)+\mathrm{O}(^{3}P)$ atomic fragments. Within the quintets, Pauli exchange limits the redistribution of valence electrons into bonding orbitals, thereby inhibiting the formation of a strong covalent bond. Among the six quintet states, $1^{5}\Pi$ is the most strongly bound, with $D_e$ = 6401~$\mathrm{cm}^{-1}$ and $R_e$ = 1.445~\AA. Its comparatively strong binding may be attributed to the $\Pi$ symmetry, which permits partial overlap between the singly occupied $p$ orbitals of carbon and oxygen, even at high spin multiplicity. By contrast, the minimum of the $1^{5}\Sigma^{+}$ state at $R_e$ = 2.432~{\AA}  is not associated with conventional covalent bonding. Instead, it corresponds to a shallow long-range well supported primarily by electrostatic and weak dispersion interactions between the high-spin atomic fragments. This interpretation is consistent with its low vibrational frequency, $\omega_e$ = 131~$\mathrm{cm}^{-1}$, and small dissociation energy, $D_e$ = 849~$\mathrm{cm}^{-1}$. Similarly, the local minimum of the $1^{5}\Sigma^{-}$ state lies more than 26000 $\mathrm{cm}^{-1}$ above the first dissociation asymptote. It should therefore be interpreted as a quasi-bound resonance state rather than as a conventionally bound electronic state. These features are consistent with the earlier quintet-state calculations of Bauschlicher \textit{et al.} \cite{Bauschlicher1993} and with the results of Bop and Gacesa \cite{Bop2025}.

The overall agreement in PECs topology between the present calculations and the available reference curves supports the state assignments used in our analysis. At the same time, the occurrence of double wells, shallow minima, and avoided crossings indicates that not every local minimum supports a conventional series of bound rovibrational levels. Accordingly, the quantities reported below for minima lying above the dissociation asymptote are interpreted as local potential parameters rather than as properties of stable bound states. The present PECs are subsequently used to determine the spectroscopic constants discussed in the next section, and, together with the corresponding electronic wavefunctions, to analyze the permanent dipole-moment functions in Section~III.C.

\subsection{Spectroscopic Constants}

Spectroscopic constants extracted from the present adiabatic PECs for the 18 electronic states of $^{12}$C$^{16}$O correlating with the first dissociation asymptote are collected in Table~\ref{tab:spectro}, together with available experimental and previous theoretical data. For bound or weakly bound states, we report the equilibrium bond length $R_e$, the electronic term energy $T_e$, the electronic term origin $T_0$, the harmonic vibrational constant $\omega_e$, the vibrational anharmonicity constant $\omega_e x_e$, the equilibrium rotational constant $B_e$, the vibration-rotation interaction constant $\alpha_e$, the equilibrium dissociation energy $D_e$, and the bond-dissociation energy $D_0$. For local minima above the asymptote, both $D_e$ and $D_0$ are negative. For predominantly repulsive states, the spectroscopic constants are not physically meaningful and are therefore omitted.

For the ground state, $1^1\Sigma^+$, we obtain $R_e=1.131$~\AA, $\omega_e=2161.04$~cm$^{-1}$, $\omega_e x_e=13.04$~cm$^{-1}$, and $B_e=1.919$~cm$^{-1}$. The absolute deviations from the experimental values compiled in \cite{HuberHerzberg1979, LeFloch2006} are 0.003~\AA, 8.77~cm$^{-1}$, 0.25~cm$^{-1}$, and 0.011~cm$^{-1}$. The present equilibrium distance $R_e$ is approximately $0.0027$~\AA\ larger than the adiabatically corrected value of $1.12827$~\AA\ reported very recently by Koput~\cite{Koput2024}. The calculated dissociation energies, $D_e=90507$~cm$^{-1}$ and $D_0=89473$~cm$^{-1}$, confirm the deep potential well associated with the strong chemical bond in CO. The latter is lower than the experimental values reported by Gong \emph{et al.} \cite{Gong2023} and K\c{e}pa \emph{et al.} \cite{Kepa2014} by 130 and 124 cm$^{-1}$, respectively.

Among the excited singlet $\Sigma^+$ states, the adiabatic $2^1\Sigma^+$ potential exhibits a pronounced double-minimum structure. The inner well is located at $R_e=1.124$~\AA\ with $T_e=87987$~cm$^{-1}$ and has a relatively large local curvature ($\omega_e=2162.72$~cm$^{-1}$) but a limited potential energy well depth, whereas the outer minimum occurs at $R_e=1.609$~\AA\ and $T_e=88898$~cm$^{-1}$ with a much softer vibrational motion ($\omega_e=598.46$~cm$^{-1}$). The shallow well depths ($D_e=2520$ and $1608$~cm$^{-1}$ for the inner and outer wells, respectively) indicate perturbations and avoided-crossing effects coming from the interaction with the higher $3^1\Sigma^+$ electronic state. For the inner and outer minima, the calculated $\omega_e$ values differ from the experimental values \cite{Eidelsberg1987} and \cite{wolk1983} by +0.97 and -52.94 cm$^{-1}$, respectively.

The low-lying excited singlet states considered here are $1^1\Sigma^-$, $1^1\Pi$, and $1^1\Delta$. For $1^1\Sigma^-$, we obtain $R_e=1.396$~\AA, $T_e=65014$~cm$^{-1}$, and $\omega_e=1090.82$~cm$^{-1}$. Relative to the experimental values \cite{HuberHerzberg1979}, $R_e$ is larger by 0.006~\AA, whereas $T_e$ and $\omega_e$ are lower by 70 and 1.40 cm$^{-1}$, respectively. The $1^1\Pi$ state is predicted at $T_e=65189$~cm$^{-1}$ with $R_e=1.237$~\AA\ and $\omega_e=1536.46$~cm$^{-1}$, while the $1^1\Delta$ state lies slightly above at $T_e=65923$~cm$^{-1}$ with $R_e=1.402$~\AA\ and $\omega_e=1065.19$~cm$^{-1}$. In the case of the $1^1\Pi$ state, the present $T_e$ and $\omega_e$ differ by 114 and 18 cm$^{-1}$, respectively, in comparison with experimental values \cite{HuberHerzberg1979}, while for the $R_e$, $\omega_ex_e$, and $B_e$ spectroscopic constants differences are only 0.002 \AA, 1.55, and 0.005 cm$^{-1}$, respectively. For $T_e$ of our $1^1\Delta$ electronic state, the difference is only 5 cm$^{-1}$ in comparison with the experimental value \cite{HuberHerzberg1979}, while for the other reported spectroscopic parameters these deviations are slightly larger than the recent theoretical results.

In the triplet states, the lowest state is $1^3\Pi$, located at $T_e=48791$~cm$^{-1}$ with $R_e=1.209$~\AA\ and $\omega_e=1753.98$~cm$^{-1}$, followed by $1^3\Sigma^+$ at $T_e=55459$~cm$^{-1}$ and $1^3\Delta$ at $T_e=60762$~cm$^{-1}$. The $1^3\Sigma^-$ state is found at $T_e=64102$~cm$^{-1}$ with $R_e=1.388$~\AA\ and $\omega_e=1116.79$~cm$^{-1}$. Overall, the predicted spectroscopic constants for these triplet states compare favorably with available measurements \cite{HuberHerzberg1979, Herzberg1950}, demonstrating the reliability of the present icMRCI(+Q) treatment. For example, in the case of the deepest triplet electronic state, $1^3\Pi$, the electronic term energy differs by 105~cm$^{-1}$ in the comparison with the experimental value \cite{HuberHerzberg1979}, while the recent theoretical result \cite{Shang2026} gives a difference of 182 ~cm$^{-1}$. Our present $\omega_e$ and $B_e$ calculated for the $1^3\Sigma^-$ state agree very well with values derived from experiment  \cite{HuberHerzberg1979}, with deviations of 0.9 and 0.005 cm$^{-1}$.

Higher excited electronic states such as $2^3\Sigma^+$, $2^1\Pi$, and $2^3\Pi$ display shallower wells and noticeable anharmonic behavior. For example, the $2^3\Sigma^+$ state has $R_e=1.090$~\AA, $T_e=84318$~cm$^{-1}$, a large vibrational constant $\omega_e=2016.55$~cm$^{-1}$, a negative value of $\omega_ex_e=-73.96$~cm$^{-1}$, and a rotational constant $B_e$ of 1.988~cm$^{-1}$. Due to the strong avoided crossing between the $1^{3}\Sigma^{+}$ and $2^{3}\Sigma^{+}$ states in the short-range repulsive region near $R=1.1$~\AA\ (Figure \ref{fig:PEC}), not reported in the previous results \cite{HuberHerzberg1979, Khalil2025}, which dramatically changes the shape of this PEC near the equilibrium position, the present spectroscopic constants differ largely in comparison with the previous experimental and theoretical results. This appears in large deviations in the values of the $\omega_e$ parameter, for which the difference between the present result and the experimental one \cite{HuberHerzberg1979} equals 183~cm$^{-1}$, while the same discrepancy between the recent theoretical value \cite{Khalil2025} and experimental measurements is 964~cm$^{-1}$. In the case of the vibrational anharmonicity constant $\omega_ex_e$, the influence of the aforementioned avoiding crossing is much more visible, because we report the negative value of this parameter ($\omega_ex_e=-73.96$~cm$^{-1}$), while Khalil \emph{et al.} \cite{Khalil2025} present a positive value equal to 228.22~cm$^{-1}$. When it comes to the $2^1\Pi$ electronic state, it possesses an irregular shape, caused interaction with the higher-lying $^1\Pi$ state, resulting in a high potential energy barrier and a quite narrow potential energy well ($\omega_e=2091$~cm$^{-1}$) with the minimum laying at $R_e=1.124$~\AA\ and 3595~cm$^{-1}$ above $\mathrm{C}(^{3}P)+\mathrm{O}(^{3}P)$ atomic asymptote. The irregular shape of the $2^1\Pi$ PEC causes the negative value of the $\omega_ex_e$ constant (-40.36~cm$^{-1}$), which in the case of Khalil \emph{et al.} \cite{Khalil2025} equals to 30.97~cm$^{-1}$.

The quintet states exhibit mixed behavior. The $1^5\Sigma^+$ state supports a very weakly bound long-range minimum at $R_e=2.432$~\AA\ with $\omega_e=131.02$~cm$^{-1}$, and $D_e=849$~cm$^{-1}$. For this electronic state, the experimental data are not available. The comparison of the present spectroscopic constants with the other theoretical results is reasonable; however, the $1^5\Sigma^+$ state is deeper by around 260~cm$^{-1}$ compared to recently reported values \cite{Rosenkrantz1992, Bauschlicher1993}. The vibrational constant is also slightly larger in comparison with these data, and the differences are 26.52 and 17.02 for the results of Rosenkrantz \emph{et al.} \cite{Rosenkrantz1992} and Bauschlicher \emph{et al.} \cite{Bauschlicher1993}, respectively. No other spectroscopic data are available to compare with the $1^5\Sigma^-$ state; we present our results for this state in Table \ref{tab:spectro}. In turn, the $1^5\Pi$ state is more compact ($R_e=1.445$~\AA, $T_e=84105$~cm$^{-1}$, $\omega_e=982.34$~cm$^{-1}$, and $D_e=6401$~cm$^{-1}$). Once again, the present PEC is deeper by 358~cm$^{-1}$ than the previously reported theoretical one \cite{Bauschlicher1993}. The $\omega_e$ constant is similar and differs by only 10~cm$^{-1}$. In contrast, the $2^5\Sigma^+$, $2^5\Pi$, and $1^5\Delta$ electronic states are found to be predominantly repulsive over the investigated range and therefore do not support stable bound vibrational levels.

The present spectroscopic constants show good overall agreement with available experimental benchmarks for the well-characterised low-lying singlet and triplet states, while also providing new reliable predictions for several poorly known or previously uncharacterized excited quintet and higher-lying states of CO.

\subsection{Permanent Dipole Moments}\label{subsection_PDM}

Permanent dipole-moment functions characterize the net molecular polarization and its dependence on internuclear
separation. Together with rovibrational wavefunctions obtained from the PECs, they can be used to calculate electric-dipole rotational and rovibrational transition intensities within individual electronic states. Here we report PDM functions for all 18 electronic states correlating with the $\mathrm{C}(^3P)+\mathrm{O}(^3P)$ limit.

For each electronic state $i$, the body-fixed PDM function was calculated  as 
\begin{equation}
\mu_i(R)=\langle \Psi_i(R)\,|\,\hat{\mu}_z\,|\,\Psi_i(R)\rangle ,
\end{equation}
using the icMRCI wavefunctions obtained from the same SA-CASSCF reference as employed for the PECs. Here $\hat{\mu}_z$ is the z-component of the total molecular electric-dipole operator. The z axis is taken along the molecular axis, from C to O. The absolute sign of $\mu(R)$ depends on the chosen axis orientation, whereas trends with $R$ and relative magnitudes for the electronic states are physically meaningful. The PDM functions were calculated on a nonuniform grid of 37 internuclear separations covering $R = 0.85$--$4.00$~\AA, with a spacing of $\Delta R = 0.05$~\AA{} between $0.85$ and $1.30$~\AA{} and $\Delta R = 0.10$~\AA{} between $1.30$ and $4.00$~\AA.

\begin{figure}
    \centering
    \includegraphics[width=\columnwidth]{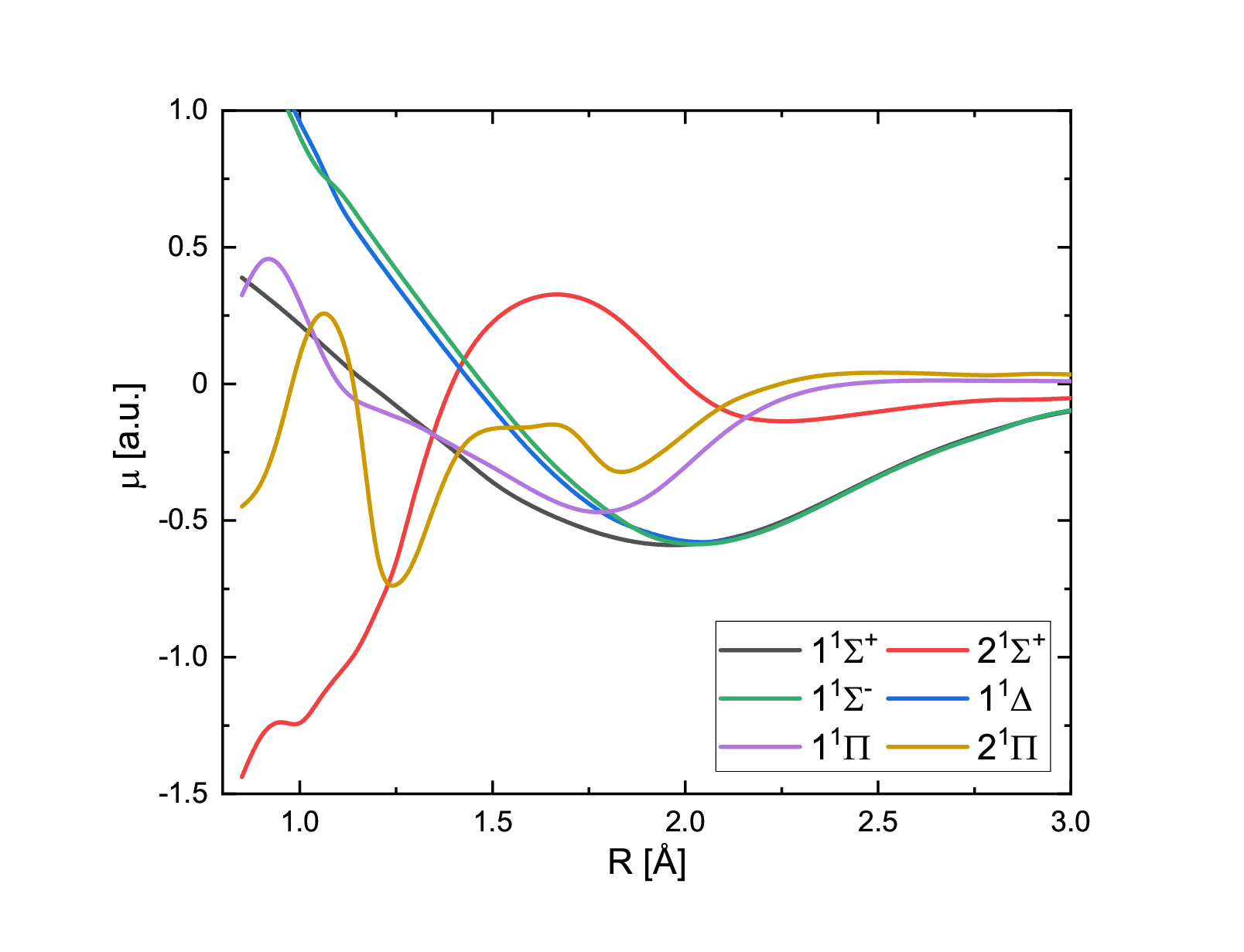}
    \caption{Permanent dipole-moment functions $\mu(R)$ for the singlet states of CO in the
    Cartesian/body-fixed representation (see text for sign convention).}
    \label{fig:PDM1}
\end{figure}

\begin{figure}
    \centering
    \includegraphics[width=\columnwidth]{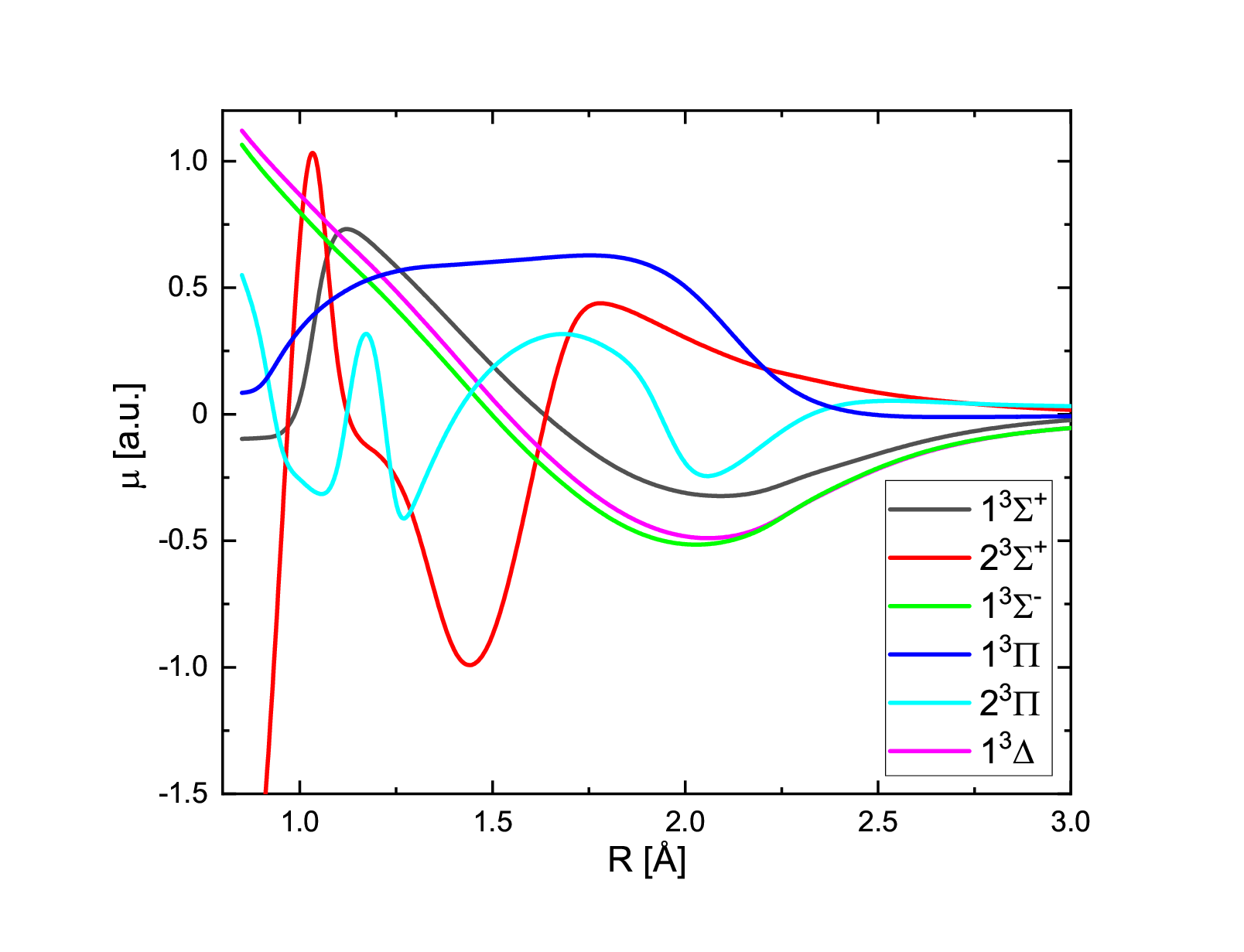}
    \caption{Permanent dipole-moment functions $\mu(R)$ for the triplet states of CO in the
    Cartesian/body-fixed representation.}
    \label{fig:PDM3}
\end{figure}

\begin{figure}
    \centering
    \includegraphics[width=\columnwidth]{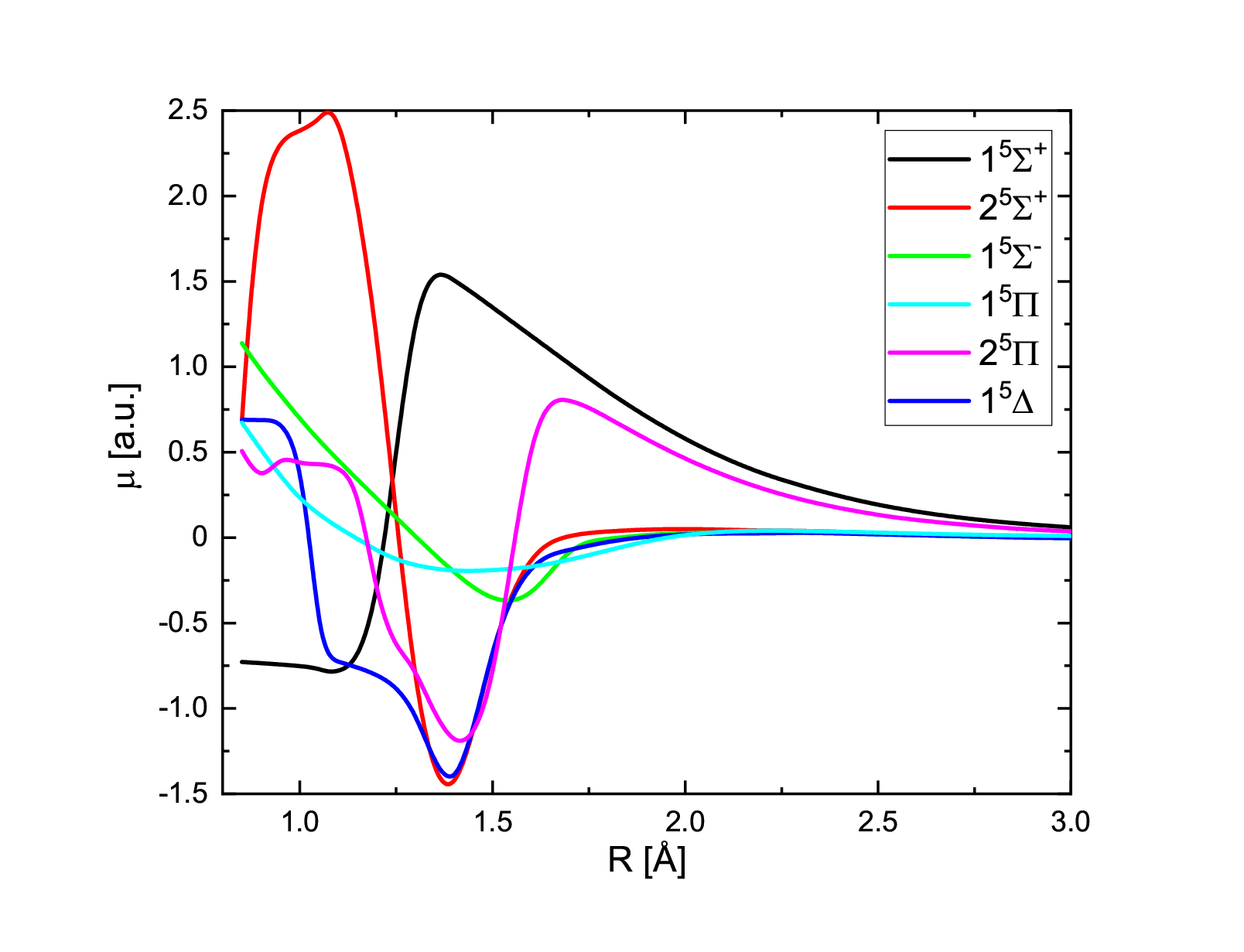}
    \caption{Permanent dipole-moment functions $\mu(R)$ for the quintet states of CO in the
    Cartesian/body-fixed representation.}
    \label{fig:PDM4}
\end{figure}

Figures~\ref{fig:PDM1}--\ref{fig:PDM4} display the resulting PDM functions for the singlet, triplet, and quintet states, grouped by $\Sigma^{\pm}$, $\Pi$, and $\Delta$ symmetries. For states correlating with the neutral-atoms asymptote, $\mu(R)$ must vanish in the separated-atoms limit ($R\rightarrow\infty$). Consequently, nonzero dipoles arise primarily at short and intermediate bond lengths where the electronic density is redistributed between the atomic centers. Extrema and sign changes can occur as the dominant configurations change.

The PDM function of the $2^{1}\Pi$ state (Fig.~\ref{fig:PDM1}) exhibits two pronounced local minima near $R=1.25$ and $1.8$~\AA. The corresponding PEC (Fig.~\ref{fig:PEC}) displays a shoulder around $R=1.3$--$1.4$~\AA\ and a broad local maximum near $R=1.75$~\AA. These features in the PEC and PDM function are associated with interaction with a higher-lying $^{1}\Pi$ state. The resulting changes in adiabatic electronic character affect both the shape of the potential and the permanent dipole moment.

For the triplet $\Sigma^{+}$ states (Fig.~\ref{fig:PDM3}), two distinct interaction regions should be distinguished. The rapid short-range PDM variations near $R=1.1$~\AA\ are associated with the avoided crossing between $1^{3}\Sigma^{+}$ and $2^{3}\Sigma^{+}$ in the repulsive region of their PECs. At larger separation, the pronounced local maximum of the $2^{3}\Sigma^{+}$ PEC near $R=1.5$~\AA\ arises from interaction with a higher-lying $^{3}\Sigma^{+}$ state. The strong variation of the $2^{3}\Sigma^{+}$ PDM in this bond-length region reflects the corresponding change in adiabatic electronic character. Near these interactions, configuration mixing can substantially alter the permanent dipole moment over a narrow interval of internuclear separation.

As described in Section~\ref{active_space_section}, the calculations included one or two additional higher-lying electronic states beyond the eighteen considered in the present investigations states, allowing these states to be examined alongside nearby higher-lying states of the same spin and spatial symmetry. These higher-lying interaction partners are not shown in Fig.~\ref{fig:PEC}, which displays only the eighteen states correlating with the $\mathrm{C}({}^{3}P)+\mathrm{O}({}^{3}P)$ asymptote.

An important aspect of the present work is the systematic characterization of the high-spin quintet states (Fig.~\ref{fig:PDM4}), which are frequently omitted in earlier studies. Although these states are predominantly repulsive in the adiabatic representation, their PDM functions remain valuable for a complete description. They are important for future extensions that account for spin-orbit and angular-momentum couplings and can play a role in quantitative assessments of coupling-driven effects on spectra and in realistic modeling of quantum dynamics.

\begin{figure}
    \centering
    \includegraphics[width=\columnwidth]{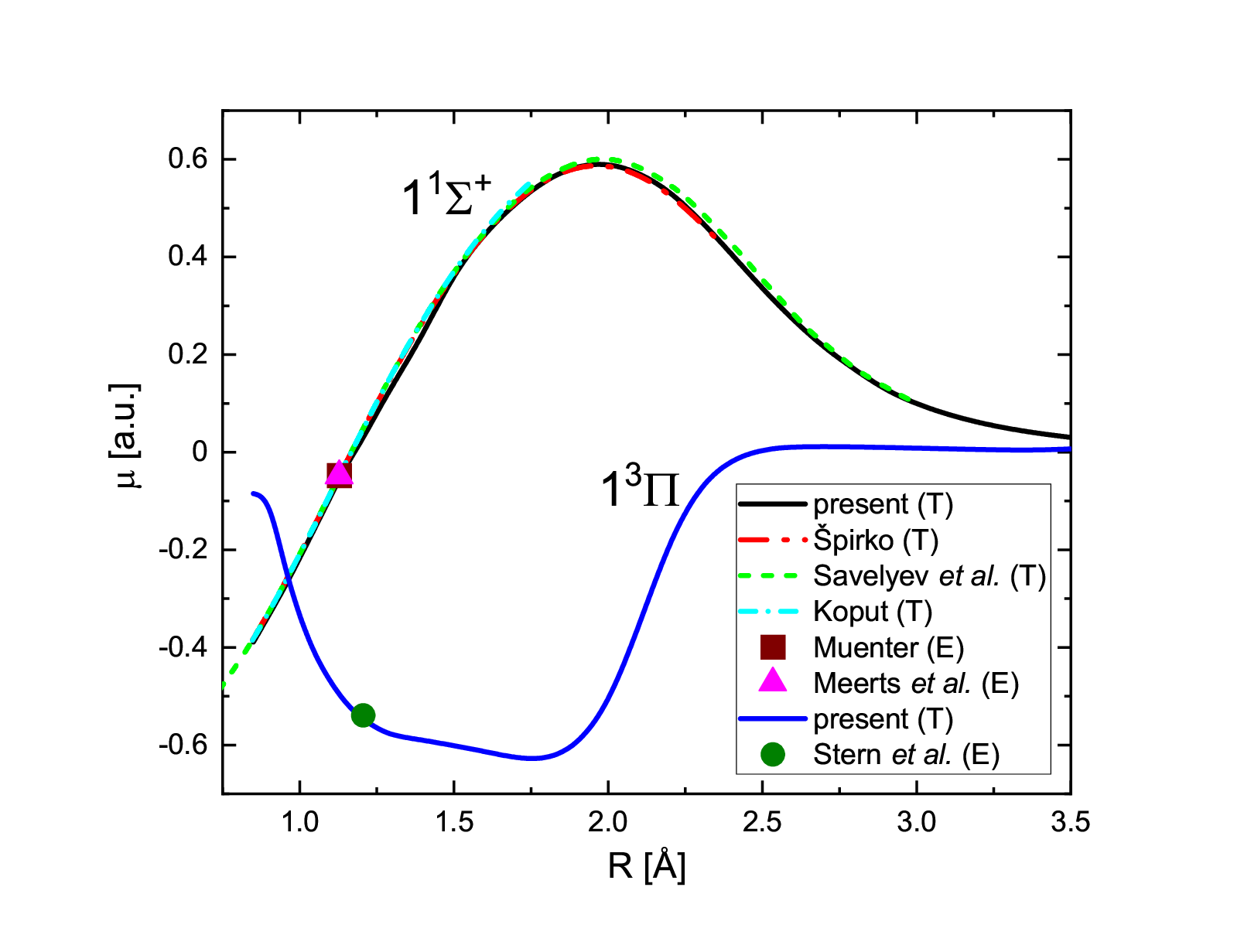}
    \caption{Permanent dipole-moment functions for the $1^{1}\Sigma^{+}$ and $1^{3}\Pi$ states of the CO molecule. The present ground-state function is compared with the ab initio curves of Koput~\cite{Koput2025} and Savelyev \textit{et al.}~\cite{Savelyev2024}, and with the semi-empirical curve of \v{S}pirko~\cite{Spirko2024}. The equilibrium dipole moments derived from experiment by Muenter~\cite{Muenter1975} and Meerts \textit{et al.}~\cite{Meerts1977} are shown at $R=1.1282$~\AA. The experimental reference value of Stern \textit{et al.}~\cite{Stern1970} is shown for the $1^{3}\Pi$ state.}
    \label{fig:EXPTH}
\end{figure}

Recent dedicated ab initio studies provide detailed benchmarks for the ground state permanent dipole moment functions of CO molecule \cite{Usov2024,Koput2024,Koput2025,Savelyev2024}. Figure~\ref{fig:EXPTH} compares the present ground state PDM function with the ab initio results of Koput \cite{Koput2025} and Savelyev \textit{et al.} \cite{Savelyev2024}, together with the semi-empirical function of \v{S}pirko \cite{Spirko2024}. The figure also includes in the comparison the equilibrium PDMs derived from experiment by Muenter \cite{Muenter1975} and Meerts \textit{et al.} \cite{Meerts1977}. At reported $R_e$ of the $1^1\Sigma^+$, the present PDM gives $\mu=-0.05026$~a.u. The equilibrium reference values of Muenter and Meerts \textit{et al.} are $-0.04808$ and $-0.04839$~a.u., respectively. The absolute differences are therefore $0.00218$ and $0.00187$~a.u. Evaluation of Koput's polynomial representation at the equilibrium distance gives $-0.04849$~a.u., differing from the present value by $0.00177$~a.u. The present dipole-moment curve agrees well with the corresponding theoretical results of Kuput \cite{Koput2025}, Savelyev \textit{et al.} \cite{Savelyev2024}, and \v{S}pirko \cite{Spirko2024}. Small deviations among the four curves may arise from differences in the electronic-structure treatment, basis-set completeness, or the fitting procedure used to reconstruct the semi-empirical curve of \v{S}pirko. Nevertheless, the close overall agreement confirms the reliability of our calculations for describing the electric properties of the CO ground state PDM function. In particular, the overall shape of the presented PDM functions in Figure~\ref{fig:EXPTH}, including the sign change at shorter bond lengths, the rise to a broad maximum near intermediate distances, and the gradual decay toward the asymptotic region, is consistently reproduced.

The change of the ground-state PDM function from negative values at short internuclear separations to positive values as $R$ increases indicates a reversal in the effective charge distribution between the carbon and oxygen centers as the bond is stretched. The broad maximum around $R \approx 2.0$~\AA\ reflects significant electronic polarization in the intermediate bonding region. At the same time, the subsequent decrease at larger separations corresponds to the progressive dissociation of the molecule into neutral atomic fragments.

For the $1^3\Pi$ state, the calculated PDM is $-0.546$~a.u. at $R = 1.200$~\AA, compared with the experimental value of $-0.539$~a.u. reported by Stern \textit{et al.} \cite{Stern1970} and plotted at $R = 1.206$~\AA{} in Figure \ref{fig:EXPTH}. This curve characterizes the broad extremum with the minimum lying at 1.8~\AA\ and the value of -0.6 a.u. Starting from 2~\AA, this PDM function quickly vanishes to zero. This means that the effective charge distribution between the carbon and oxygen centers is different for the higher-spin triplet electronic state compared to the singlet ground state.

Overall, the present calculations provide an accurate representation of permanent dipole moment functions for the CO molecule over a wide range of internuclear distances, supporting their use in rovibrational intensity calculations and molecular spectroscopy.

\begin{table*}[ht!]
\centering
\caption{Spectroscopic constants for all 18 electronic states of CO that correlate with the first atomic dissociation asymptote, obtained with the icMRCI(+Q) method.}
\begin{footnotesize}
\begin{tabular}{l c c c c c c c c c c}
\hline\hline
Ref. & State & $R_e$ [\AA] & $T_e$ [cm$^{-1}$] & $T_0$ [cm$^{-1}$] & $\omega_e$ [cm$^{-1}$] & $\omega_e x_e$ [cm$^{-1}$] & $B_e$ [cm$^{-1}$] & $\alpha_e$ [cm$^{-1}$] & $D_e$ [cm$^{-1}$] & $D_0$ [cm$^{-1}$] \\
\hline
This work &  $1^1\Sigma^+$ & 1.131 & 0 & 0 & 2161.04$\pm$1.56 & 13.04$\pm$0.08 & 1.919$\pm$0.0005 & 0.0171$\pm$0.0001 & 90507 & 89473 \\
Expt$^{j}$ & &  &  &  &  &  &  &  &  &  89603  \\
Expt$^{k}$ & &  &  &  &  &  &  &  &  &  89597  \\
Expt$^{a}$ & & 1.128 & 0 &  & 2169.81 & 13.29 & 1.930 &  &  &  \\
Expt$^{b}$ & & 1.128 & 0 &  & 2169.76 & 13.29 & 1.930 &  &  &   \\
Theory$^{h}$ & & 1.128 &  &  &  &  & 1.923 &  &  &   \\
Theory$^{d}$ & & 1.13 & 0 &  & 2172.18 & 12.78 & 1.970 &  &  &  \\
Theory$^{i}$ & & 1.14 & 0 &  & 2125.45 & 12.29 & 1.90 &  &  &  \\

This work & $2^1\Sigma^+$ (1$^{st}$ min) & 1.124 & 87987 & 88009 & 2162.72& 44.04 & 1.970$\pm$0.0302 &0.0552$\pm$0.0177 &2520 & 1464 \\
Expt$^{a}$ & & 1.120 & 86945 &  & 2112.70 & 15.22 & 1.960  \\
Expt$^{e}$ & & 1.119 & 86927 &  & 2161.75 & 39.84 &  \\
Theory$^{d}$ & & 1.12 & 86945 &  & 2135.32 & 13.52 & 1.960 &  &  &  \\
Theory$^{i}$ & & 1.13 & 87964 &  & 2152.16 & 62.35 & 1.92 &  &  &  \\

This work & $2^1\Sigma^+$ (2$^{nd}$ min) & 1.609 & 88898 & 88218 & 598.46$\pm$20.40& 16.69$\pm$3.96 & 0.940$\pm$0.012 &-0.0520$\pm$0.0117 &1608 & 1255 \\
Expt$^{f}$ & & 1.583 & 89438 &  & 651.40$\pm$5.70 & 20.40$\pm$2.50 & 0.981&0.0270\\
Theory$^{i}$ & & 1.65 & 87338 &  & 551.77 & 30.68 & 0.90 &  &  &  \\

This work & $1^3\Sigma^+$ & 1.357 & 55459 & 55050 & 1237.42$\pm$2.25 & 13.73$\pm$0.33 & 1.334$\pm$0.001 & 0.0167$\pm$0.0002 & 35047 & 34423 \\
Expt$^{a}$ & & 1.352 & 55825 &  & 1228.60 & 10.47 & 1.345  \\
Expt$^{c}$ & & 1.351 & 55823 &   & 1230.65 & 11.01 & 1.345 \\
Theory$^{d}$ & & 1.36 & 48868 &  & 1250.79 & 10.18 & 1.720 &  &  &  \\
Theory$^{i}$ & & 1.36 & 54371 &  & 1177.20 & 29.61 & 1.33 &  &  &  \\

This work &  $2^3\Sigma^+$ & 1.090 & 84318 & 84434 & 2016.55$\pm$26.03 & -73.96$\pm$6.06 & 1.988$\pm$0.012 & 0.0523$\pm$0.0062 & 6188 & 5039 \\
Expt$^{a}$ & & 1.113 & 83814 &  &  2199.30 &  & 1.986  \\
Theory$^{i}$ & & 1.11 & 82912 &  & 3163.71 & 228.22 & 1.98 &  &  &  \\

This work & $1^5\Sigma^+$ & 2.432 & 89658 & 88688 & 131.02$\pm$0.06 & 5.83$\pm$0.02 & 0.415$\pm$0.0003 & 0.0230$\pm$0.0002 & 849 & 784 \\
Theory$^{l}$ &  & 2.571  &  &   &104.5  &  &  &  & 587 &    \\
Theory$^{m}$ &  & 2.519  &  &   &114  &  &  &  & 594 &    \\

This work &  $2^5\Sigma^+$ & Repulsive &  &  &  &  &  \\

This work & $1^1\Sigma^-$ & 1.396 & 65014 & 64517 & 1090.82$\pm$1.66 & 12.84$\pm$0.30 & 1.262$\pm$0.001 & 0.0192$\pm$0.0003 & 25492 & 24955\\
Expt$^{a}$ & & 1.39 & 65084 &  & 1092.22 & 10.70 & 1.271  \\
Theory$^{d}$ & & 1.39 & 64869 &  & 1096.30 & 10.78 & 1.26 &  &  &  \\
Theory$^{i}$ & & 1.40 & 64366 &  & 1041.22 & 53.63 & 1.25 &  &  &  \\

This work & $1^3\Sigma^-$ & 1.388 & 64102 & 63624 & 1116.79$\pm$1.16 & 10.46$\pm$0.17 & 1.279$\pm$0.001 & 0.0179$\pm$0.0002 & 26405 & 25849 \\
Expt$^{a}$ & & 1.384 & 64230 &  & 1117.72 & 10.69 & 1.284  \\
Expt$^{c}$ & & 1.393 & 65340 &  & 1093.99 & 9.58 & 1.266  \\
Theory$^{d}$ & & 1.38 & 64109 &  & 1123.67 & 10.69 & 1.370 &  &  &  \\
Theory$^{i}$ & & 1.40 & 63375 &  & 1027.51 & 9.66 & 1.26 &  &  &  \\

This work & $1^5\Sigma^-$ & 1.371 & 116948 & 116486 &1194.64 & 31.89 & 1.430$\pm$0.141 & 0.1714$\pm$0.0824 & -26441 & -27012 \\

This work & $1^1\Pi$ & 1.237 & 65189 & 64910 & 1536.46$\pm$1.59 & 20.95$\pm$0.25 & 1.607$\pm$0.001 & 0.0237$\pm$0.0002 & 25317 & 24563 \\
Expt$^{a}$ & & 1.235 & 65075 &  & 1518.24 & 19.40 & 1.612  \\
Theory$^{d}$ & & 1.24 & 65158 &  & 1516.77 & 17.76 & 1.630 &  &  &  \\
Theory$^{i}$ & & 1.24 & 65036 &  & 1497.22 & 15.12 & 1.59 &  &  &  \\

This work & $2^1\Pi$ & 1.124 & 94102 & 94144 & 2090.89$\pm$11.00 & -40.36$\pm$5.14 & 1.895$\pm$0.037 & -0.0767$\pm$0.0352 & -3595 & -4670 \\
Expt$^{a}$ & & 1.115 & 92903 &  & 2153.80 &  & 1.977  \\
Theory$^{i}$ & & 1.13 & 94408 &  & 2121.27 & 30.97 & 1.92 &  &  &  \\

This work & $1^3\Pi$ & 1.209 & 48791 & 48618 & 1753.98$\pm$5.83 & 14.57$\pm$0.41 & 1.687$\pm$0.002 & 0.0200$\pm$0.0006 & 41715 & 40855 \\
Expt$^{a}$ & & 1.21 & 48686  &  & 1743.41 & 14.36 & 1.690 \\
Theory$^{d}$ & & 1.21 & 48868 &  & 1750.70 & 15.10 & 1.720 &  &  &  \\

This work & $2^3\Pi$  & 1.382 & 91330 & 90682 & 710.84$\pm$51.29 & -65.27$\pm$14.90 & 1.179$\pm$0.152 & -0.1029$\pm$0.0784 & -823 & -1209 \\
Expt$^{a}$ & & 1.127 & 93158 &  &  &  & 1.9771  \\
Expt$^{g}$ & & 1.379 & 91012 &  &805.12  & -2.85  & 1.293  \\

This work & $1^5\Pi$ & 1.445 & 84105 & 83575 & 982.34$\pm$6.15 & 16.19$\pm$2.05 & 1.171$\pm$0.004 & 0.0255$\pm$0.0031 & 6401 & 5898 \\
Theory$^{m}$ &  & 1.458  &  &   & 972  &  &  &  & 6043  &    \\

This work & $2^5\Pi$ & Repulsive &  &  &  &  &  \\

This work & $1^1\Delta$ & 1.402 & 65923 & 65408 & 1065.19$\pm$3.47 & 12.13$\pm$0.45 & 1.252$\pm$0.002 & 0.0192$\pm$0.0005 & 24583 & 24065 \\
Expt$^{a}$ & & 1.399 & 65928 &  & 1094.00 & 10.20 & 1.257  \\
Theory$^{d}$ & & 1.40 & 65426 &  & 1099.35 & 11.97 & 1.340 &  &  &  \\

This work & $1^3\Delta$ & 1.374 & 60762 & 60318 & 1170.74$\pm$2.81 & 11.60$\pm$0.31 & 1.297$\pm$0.001 & 0.0154$\pm$0.0001 & 29744 & 29155 \\
Expt$^{a}$ & & 1.370 & 61120 &  & 1171.94 & 10.64 & 1.311  \\
Theory$^{d}$ & & 1.37 & 60959 &  & 1188.60 & 10.05 & 1.410 &  &  &  \\
Theory$^{i}$ & & 1.38 & 60226 &  & 1075.77 & 28.45 & 1.30 &  &  &  \\

This work & $1^5\Delta$ & Repulsive &  &  &  &  &  \\

\hline\hline
\end{tabular}
\end{footnotesize}
\vspace{1ex}
\raggedright
\textit{Note.} $^{a}$Huber \& Herzberg\cite{HuberHerzberg1979}, $^{b}$Le Floch\cite{LeFloch2006},  $^{c}$Herzberg \& Hugo\cite{Herzberg1950}, $^{d}$Shang et al.\cite{Shang2026}, $^{e}$Eidelsberg et al.\cite{Eidelsberg1987}, $^{f}$Wolk \& Rich\cite{wolk1983}, $^{g}$Berden et al.\cite{Berden1997},
$^{h}$Koput\cite{Koput2024}, $^{i}$Khalil\cite{Khalil2025},  $^{j}$Gong et al.\cite{Gong2023},  $^{k}$Kepa et al.\cite{Kepa2014}, $^{l}$Rosenkrantz et al.\cite{Rosenkrantz1992} and  $^{m}$Bauschlicher et al.\cite{Bauschlicher1993} \label{tab:spectro}
\end{table*}

\section{Uncertainty estimates}
\label{sec:uncertainty}

In this section, we summarize the dominant sources of uncertainty affecting the reported PECs and the spectroscopic parameters (SP) derived from these curves. The main contributions are (i) basis-set incompleteness, (ii) residual electron-correlation and active-space effects, and (iii) enhanced sensitivity in regions of adiabatic state mixing (e.g., in avoided-crossing regions). The discussion below first focuses on the analysis of CBS limit calculations, which provide the most direct quantitative estimate of uncertainty in the present set of theoretical results.
\begin{figure}
    \centering
    \includegraphics[width=\columnwidth]{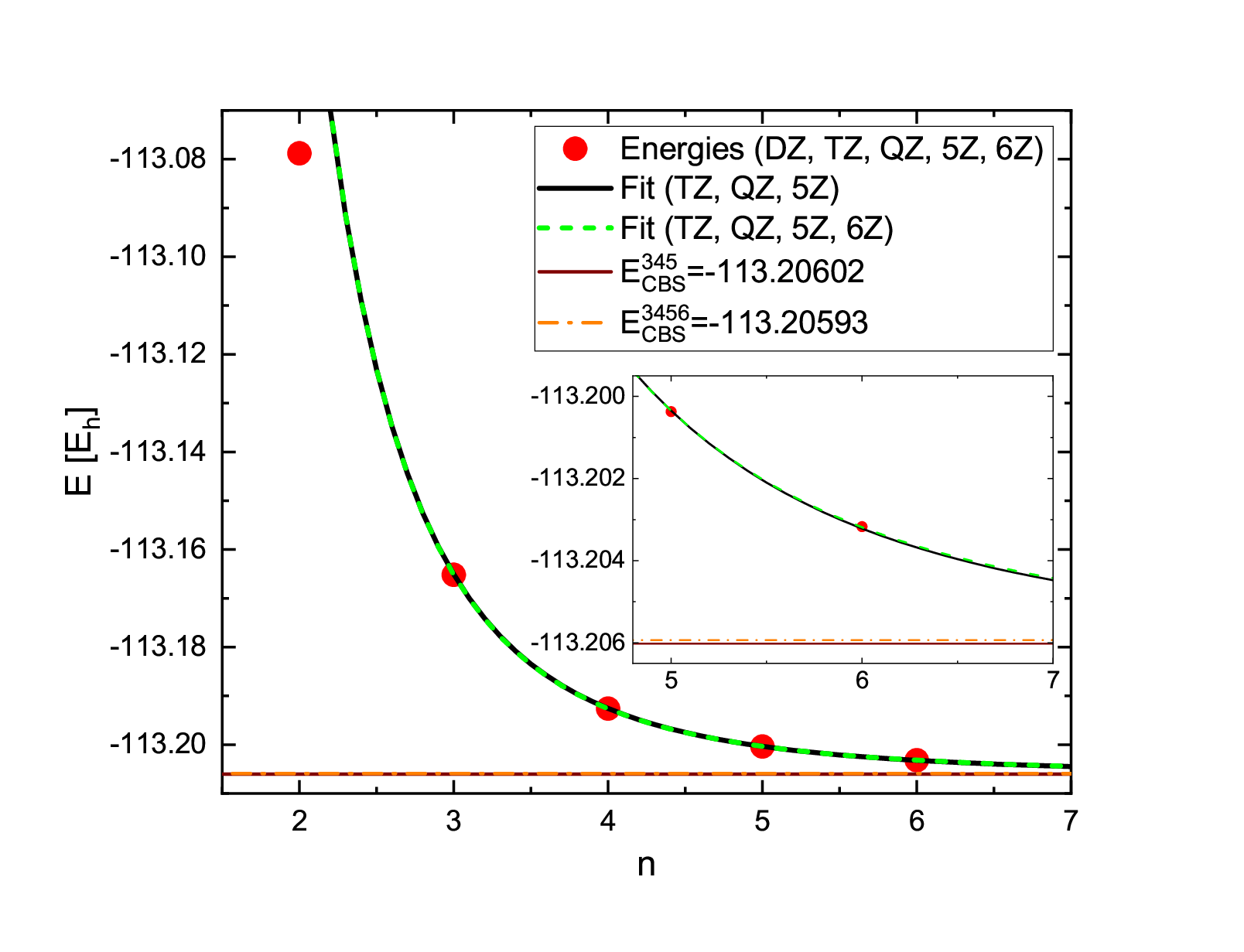}
    \caption{The results of CBS limit extrapolation (E$^{345}_{CBS}$ and E$^{3456}_{CBS}$) obtained by fitting the inverse-power model in Eq. (1) to the CO ground-state energies calculated at the equilibrium point (red circles) with aug-cc-pV$n$Z basis sets for cardinal numbers $n$=3,4,5 (black solid line) and $n$=3,4,5,6 (dashed green line). The aug-cc-pVDZ basis set was excluded from fitting. The small difference between E$^{345}_{\mathrm{CBS}}$ and E$^{3456}_{\mathrm{CBS}}$ is visible in the inset.}
    \label{fig:CBS}
\end{figure}

To assess the convergence of the investigated $^{12}$C$^{16}$O ground-state potential energy with the size of the computational basis set, the energies of the 1$^1\Sigma^+$ electronic state at the equilibrium distance R$_e$ were computed using the augmented correlation-consistent basis sets aug-cc-pV$n$Z and extrapolated toward the CBS limit using the inverse-power model introduced in Eq.~(1). Figure~\ref{fig:CBS} illustrates this extrapolation and the resulting CBS estimate. The extrapolation was carried out using the basis-set sequence described in Section~II.C.

To assess the residual basis-set incompleteness uncertainty of the extrapolated CBS limit, we compared independent E$^{345}_{\mathrm{CBS}}$ and E$^{3456}_{\mathrm{CBS}}$ estimates obtained from two different extrapolation datasets based on the cardinal-number sequences $n$=3,4,5 and $n$=3,4,5,6. Rather than adopting the difference between these estimates as the uncertainty itself, we defined the best CBS value as the arithmetic mean of the two independent extrapolations,
\begin{equation}
E_{\mathrm{CBS}}=\frac{E_{\mathrm{CBS}}^{(345)}+E_{\mathrm{CBS}}^{(3456)}}{2},
\end{equation}
while the associated uncertainty was estimated as one-half of their difference,
\begin{equation}
\Delta E_{\mathrm{CBS}} = \frac{1}{2}\left|E_{\mathrm{CBS}}^{(345)}-E_{\mathrm{CBS}}^{(3456)}\right|.
\end{equation}
This approach provides a symmetric uncertainty interval around the best CBS estimate and avoids overestimating the residual basis-set error by assigning the entire spread between the two extrapolations to the uncertainty. For the present calculations, the E$^{345}_{\mathrm{CBS}}$ and E$^{3456}_{\mathrm{CBS}}$ limits differ by only 18 cm$^{-1}$, resulting in an estimated extrapolation spread of $\pm$9 cm$^{-1}$. The small magnitude of this uncertainty demonstrates the excellent convergence of the icMRCI(+Q) energies with respect to the basis-set size and indicates that the remaining basis-set incompleteness error (BSIE) is negligible compared with the overall accuracy required for the present study.
 
The uncertainty of the CBS extrapolation was further assessed using the stochastic random-walk approach proposed by Lang, Przybytek, and Lesiuk \cite{Lang2025}. In this procedure, a sequence of extrapolated CBS estimates from consecutive basis-set pairs is used to generate an ensemble of random walks that represent possible continuations of the extrapolation sequence beyond the largest explicitly calculated basis set. The resulting distribution of asymptotic values provides a statistically meaningful estimate of the residual uncertainty in CBS extrapolation.

In our approach, the icMRCI(+Q) molecular energies calculated at the equilibrium distance R$_e$  of the CO ground state with the aug-cc-pV$n$Z ($n$ = 3,4,5,6) basis sets were used as input data. The two-point Truhlar/Helgaker-type extrapolation formula \cite{Helgaker1997, HALKIER1998, TRUHLAR1998} 
\begin{equation}
E_{\mathrm{CBS}}(n/m)=\frac{m^pE(m)-n^pE(n)}{m^p-n^p}
\end{equation}
was employed to generate the initial CBS estimates from consecutive basis-set pairs. The random-walk simulations were carried out using 10$^{5}$--10$^{8}$ trajectories as shown in Table \ref{tab:cbs_rw}. The uncertainty intervals were extracted from the distribution of the converged random-walk values and are reported as 1$\sigma$, 2$\sigma$, and 3$\sigma$ confidence intervals (CIs) corresponding to 68.27\%, 95.45\%, and 99.73\% of the generated distribution, respectively.

The sensitivity of the random-walk uncertainty estimate to the extrapolation exponent $p$ was examined by varying $ p$ from 3.0 to 5.0. The value $p$=4 was chosen because it yielded mutually consistent CBS estimates from the QZ/5Z and 5Z/6Z basis-set pairs. This indicates that, for the present icMRCI(+Q) molecular energies, the effective basis-set convergence is better represented by an n$^{-4}$-type behavior than by the standard n$^{-3}$ correlation-energy asymptote. This further confirms the findings from the inverse-power model curve fit described above, in which $\alpha_{345} = 3.873$ and $\alpha_{3456} = 3.890$ are very close to 4. The E$_{\mathrm{CBS}}$ limit obtained in this procedure is -113.20597 E$_\mathrm{h}$ and is near coincidence to the CBS random-walk value, -113.20578 E$_\mathrm{h}$. This agreement confirms that the random-walk procedure primarily quantifies the uncertainty of the extrapolated value rather than shifting the CBS estimate itself. The  1$\sigma$ standard deviation of the random-walk distribution was 4.4 cm$^{-1}$, while the 2$\sigma$ and 3$\sigma$ confidence intervals were 9.9 and 16.2 cm$^{-1}$, respectively. All these values are presented in Table \ref{tab:cbs_rw}. Thus, the final CBS equilibrium energy of the ground state may be reported as E$_{CBS}^{RW}$ = -113.205781 $\pm$ 0.000045 E$_\mathrm{h}$, where the uncertainty corresponds to an approximately 95\% confidence interval obtained from the random-walk analysis.

The very small CBS uncertainties obtained for both approaches, together with the near coincidence of the extrapolated values, demonstrate that the icMRCI(+Q) energies are well converged with respect to the basis-set size. The remaining BSIE is therefore expected to be minor on the energy scale relevant to the present study.

\begin{table*}[ht!]
\centering
\caption{Complete-basis-set (CBS) energies and associated uncertainties
estimated using the random-walk approach following
Lang \textit{et al.}~\cite{Lang2025}. Reported uncertainties correspond to
$1\sigma$, $2\sigma$ and $3\sigma$ confidence intervals. All energies are given in Hartree.}
\label{tab:cbs_rw}
\begin{tabular}{cccccccc}
\hline\hline
Basis sets included & $N_{\mathrm{walk}}$ & $E_{\mathrm{CBS}}$ & CI 2.5\% & CI 97.5\% & $1\sigma$ & $2\sigma$ & $3\sigma$ \\
\hline
$\{$TZ, QZ$\}$ $\{$QZ, 5Z$\}$
& $1\times10^{5}$  & -113.205749 & -113.206381 & -113.205118 & $\pm$0.000288 & $\pm$0.000647 & $\pm$0.001092 \\
$\{$TZ, QZ$\}$ $\{$QZ, 5Z$\}$
& $1\times10^{6}$  & -113.205751 & -113.206387 & -113.205116 & $\pm$0.000289 & $\pm$0.000651 & $\pm$0.001087 \\
$\{$TZ, QZ$\}$ $\{$QZ, 5Z$\}$
& $1\times10^{7}$  & -113.205751 & -113.206387 & -113.205116 & $\pm$0.000290 & $\pm$0.000652 & $\pm$0.001086 \\
$\{$TZ, QZ$\}$ $\{$QZ, 5Z$\}$
& $1\times10^{8}$  & -113.205751 & -113.206386 & -113.205115 & $\pm$0.000290 & $\pm$0.000652 & $\pm$0.001086 \\
\hline
$\{$QZ, 5Z$\}$ $\{$5Z, 6Z$\}$
& $1\times10^{5}$  & -113.205781 & -113.205824 & -113.205737 & $\pm$0.000020 & $\pm$0.000044 & $\pm$0.000075 \\
$\{$QZ, 5Z$\}$ $\{$5Z, 6Z$\}$
& $1\times10^{6}$  & -113.205781 & -113.205824 & -113.205737 & $\pm$0.000020 & $\pm$0.000045 & $\pm$0.000074 \\
$\{$QZ, 5Z$\}$ $\{$5Z, 6Z$\}$
& $1\times10^{7}$  & -113.205781 & -113.205824 & -113.205737 & $\pm$0.000020 & $\pm$0.000044 & $\pm$0.000074 \\
$\{$QZ, 5Z$\}$ $\{$5Z, 6Z$\}$
& $1\times10^{8}$  & -113.205781 & -113.205824 & -113.205737 & $\pm$0.000020 & $\pm$0.000045 & $\pm$0.000074  \\
\hline\hline
\end{tabular}
\end{table*}

The above CBS limit analyses were performed for the potential energy at a single point on the $1\,^1\Sigma^+$ ground-state PEC, evaluated at R$_e$. We also calculated the CBS limit along with the 95\% CI for the entire PEC of the ground state, as shown in Figure \ref{fig:CCSDT}. The 1$^1\Sigma^+$ ground-state PEC calculated with the aug-cc-pV6Z basis set lies close to the CBS potential-energy curve, approaching the 95\% confidence interval at intermediate distances. The BSIE at R$_e$ obtained from 5Z/6Z results does not exceed 10 cm$^{-1}$ at the 2$\sigma$ confidence level. The median pointwise 95\% uncertainty half-width obtained from the random-walk distributions along the ground state PEC is 141 cm$^{-1}$. Results such as PECs, T$_e$, T$_0$, D$_e$, and D$_0$ presented in this work were obtained using the aug-cc-pV5Z basis set, which allows us to assume that the conservative error in the calculated potential energies at the 95\% confidence level does not exceed 141 cm$^{-1}$. Because the same basis set and electronic-structure protocol were used throughout, we use this value as an indicative uncertainty scale for the other PECs by analogy with the ground-state benchmark.

Based on this analysis, our bond-dissociation energy for the 1$^1\Sigma^+$ ground state CO is  D$_0$ = 89473 $\pm$ 141 cm$^{-1}$. The central value is lower than the experimental values of 89603 and 89597 cm$^{-1}$ reported by Gong \emph{et al.} \cite{Gong2023} and K\c{e}pa \emph{et al.} \cite{Kepa2014} by 130 and 124 cm$^{-1}$, respectively. On the other hand, if we adopt the 2$\sigma$ uncertainty of $143~\mathrm{cm^{-1}}$ obtained from the 5Z-based random-walk CBS extrapolation at the single R$_e$ point, the error bar estimation for our bond-dissociation energy of the 1$^1\Sigma^+$ ground state is consistent with the above findings. Thus, we report a percentage error (PE) of $0.16\%$ for the bond-dissociation energy of the ground state of the CO molecule. For $1\sigma$ CI, the uncertainty decreases to $64~\mathrm{cm^{-1}}$ and $0.07\%$. When the 6Z basis set is included in the stochastic random-walk extrapolation, the uncertainty decreases to $9.9~\mathrm{cm^{-1}}$ ($0.01\%$) at $2\sigma$ and $4.4~\mathrm{cm^{-1}}$ ($0.005\%$) at $1\sigma$, demonstrating a substantial improvement in the reliability of the CBS limit upon extension of the basis-set hierarchy.

Uncertainties in reported spectroscopic constants presented in Table \ref{tab:spectro} arise from both the electronic-structure calculations of the underlying PECs and from the numerical fitting procedure. The transition term energies (T$_e$ and T$_0$) and dissociation energies (D$_e$ and D$_0$) are mostly affected by uncertainties associated with the BSIE discussed above. The errors in the rovibrational SPs ($\omega_e$, $\omega_e x_e$, $B_e$, and $\alpha_e$) are mainly connected with the fitting approach. For example, the vibrational constants of the CO molecule ground state calculated using aug-cc-pVQZ, aug-cc-pV5Z, and aug-cc-pV6Z basis sets are consistent and equal to 2159.05, 2161.04, 2157.22 $\mathrm{cm^{-1}}$, respectively. The calculated uncertainties for these parameters are 2.06, 1.56, 1.25 $\mathrm{cm^{-1}}$, respectively. Based on the fitting procedure described in Section II.D, we estimated error bars on the 1$\sigma$ level for rovibrational SPs, which are shown in Table \ref{tab:spectro}. 

Additionally, we performed a general analysis of spectroscopic parameter residuals, defined as the deviations between the reported SP values presented in Table \ref{tab:spectro} and the constants derived from experiments. Based on these residuals, the mean absolute errors (MAE) and the root mean squared errors (RMSE) were calculated. We chose RMSE for further analysis as a more conservative error measure. For the equilibrium bond length R$_e$, this error is only 0.009 \AA, which, in the case of the 1$^1\Sigma^+$ ground state, means the PE is 0.80\%. In turn, the RMSE of the electronic term energy T$_e$ is 422 $\mathrm{cm^{-1}}$, so for the lowest-lying excited electronic state 1$^3\Pi$ it gives the percentage error of 0.87\%, while, for example, in the case of 1$^1\Sigma^-$ it is 0.65\%. When it comes to the rovibrational constants, the RMSEs calculated for $\omega_e$ and B$_e$ are 36.55 and 0.037 $\mathrm{cm^{-1}}$, respectively. The PEs are slightly larger in these cases. The $\omega_e$ PE of the 1$^1\Sigma^+$ ground state is 1.69\%, while for the 1$^1\Pi$ it is 2.38\%. In turn, the percentage error of the rotational constant B$_e$ calculated for the ground state of the CO molecule is equal to 1.95\%, and for 1$^3\Sigma^-$ it is 2.92\%. The above analysis confirms relatively small uncertainties of the reported spectroscopic constants and good agreement with the experimental results. 

\section{Conclusions}\label{sec:conclusions}

This work presents a comprehensive and internally consistent set of eighteen adiabatic potential-energy curves for carbon monoxide correlating with the first dissociation limit, C($^3$P) + O($^3$P), computed at the SA-CASSCF/icMRCI(+Q) level of theory with the aug-cc-pV5Z basis set, supplemented by basis-set convergence tests and CBS extrapolation for the ground state. A CAS(10,11) reference and a consistent frozen-core icMRCI(+Q) treatment were used for all three spin multiplicities and all spatial symmetries considered. The computed spectroscopic constants for the ground state 1$^1\Sigma^+$ and low-lying excited states (1$^1\Sigma^-$, 1$^1\Pi$, 1$^1\Delta$, 1$^3\Sigma^+$, 1$^3\Sigma^-$, 1$^3\Pi$, and 1$^3\Delta$) are in good agreement with experimental measurements, validating the accuracy and reliability of our computational approach. Several previously unreported features have been identified in the potential-energy curves, including the strong avoided crossing between the 1$^{\mathrm{st}}$ and 2$^{\mathrm{nd}}$ $^3\Sigma^+$ electronic states that occurred near the equilibrium position of the $2^3\Sigma^+$, causing the large deviations in the values of spectroscopic parameters. The systematic inclusion and characterization of the high-spin quintet states (1,2$^5\Sigma^+$, 1$^5\Sigma^-$, 1,2$^5\Pi$, and 1$^5\Delta$), which are often omitted in prior studies, represents a significant contribution to the understanding of CO's complete electronic structure. Although these quintet states are largely repulsive or only weakly bound in the adiabatic representation, their characterization is essential for future investigations involving spin-orbit and angular-momentum couplings. In addition to the potential-energy curves, we have computed permanent dipole-moment functions for all eighteen states, revealing a complex dependence on internuclear separation and sensitivity to changes in the adiabatic wavefunction composition. The resulting PDM functions allow the molecular polarity of the singlet, triplet, and quintet states to be compared within a common computational framework. The extrema and sign changes of these functions describe state-dependent variations in the magnitude and direction of the net molecular polarisation as the bond is stretched. Together with the PECs and the spectroscopic constants obtained for bound states and characterised local minima, these results provide a consistent reference for the spectroscopic and dipole properties of the eighteen electronic states of the CO molecule.

The present dataset complements recent theoretical studies directed primarily toward the rovibronic spectroscopy of selected electronic states of CO. Zhang et al. calculated rovibronic spectra for eight singlet and triplet states using hybrid RKR/\textit{ab initio} potential-energy curves that incorporate spin--orbit and electronic angular-momentum couplings. Their study focuses on observable transition spectra, line lists, and high-temperature Boltzmann-averaged absorption cross sections for UV and VUV radiative-transfer applications. A related investigation by Khalil et al. considered the $1^{1}\Pi$--$1^{1}\Sigma^{+}$, $2^{1}\Sigma^{+}$--$1^{1}\Sigma^{+}$, $3^{1}\Sigma^{+}$--$1^{1}\Sigma^{+}$, and $2^{1}\Pi$--$1^{1}\Sigma^{+}$ band systems using \textit{ab initio} potential-energy curves and permanent and transition dipole-moment functions. Particular emphasis was placed on the diabatic treatment of interacting excited states and on the calculation of synthetic spectra for comparison with laboratory and astronomical observations. Whereas these studies were designed primarily to describe selected transition systems and their observable rovibronic spectra, the present work provides a systematic, spin-free characterization of the complete set of 18 electronic states correlating with the first atomic asymptote, including adiabatic potential-energy curves, spectroscopic constants, and permanent dipole-moment functions. It also includes all six quintet states, which were not considered in either of the aforementioned studies.

The resulting dataset provides a rigorous benchmark for electronic-structure theory and a comprehensive reference for the modeling, analysis, and assignment of molecular spectra in both laboratory and astrophysical environments. It also establishes a foundation for future studies incorporating spin--orbit and nonadiabatic couplings, and angular momentum matrix elements, including the construction of diabatic representations and quantum-dynamical simulations of photodissociation, predissociation, collision-induced energy-transfer processes, and quantum reactive scattering. The complete set of potential-energy curves and permanent dipole-moment functions presented here therefore advances the theoretical description of the electronic structure of the $^{12}$C$^{16}$O molecule. It provides a valuable resource for research in molecular spectroscopy, astrophysics, atmospheric chemistry, and fundamental quantum chemistry.

\section*{Acknowledgements}
We acknowledge support by the COST Action CA21101 ''Confined Molecular Systems: From a New Generation of Materials to the Stars''. We acknowledge partial support from the Polish Ministry of Science and Higher Education within the IDUB project. Calculations have been carried out using resources provided by the Centre of Informatics Tricity Academic Supercomputer \& Network and Wroc\l aw Centre for Networking and Supercomputing.

We sincerely appreciate Professor Hamid Berriche and Dr. Eng. Tymon Kilich for their valuable discussions during the preparation of the manuscript.

\section*{Author Contributions}
\noindent P.J. and B.A.I. contributed to the conceptualization, data curation, formal analysis, investigation, methodology, software creation, validation, visualization, and writing the original draft; N.B. contributed to the conceptualization, formal analysis, investigation, methodology, validation, writing – review $\&$ editing; S.R and R.H contributed to the investigation, methodology, validation, and writing – review $\&$ editing; J.E.S contributed to the conceptualization, formal analysis, funding acquisition, investigation, methodology, project administration, resources, supervision, validation, writing the original draft, and writing – review $\&$ editing.

\section*{Data availability}
\noindent All data reported in the current article are publicly available in the Bridge of Knowledge Open Data Repository \cite{pjasik2026}.

\section*{Appendix}
\noindent Correspondence between numerical and conventional labels of the CO electronic states. \\

\begin{tabular}{ll}
\hline
Numerical label & Conventional label \\
\hline
$1^{1}\Sigma^{+}$ & $X^{1}\Sigma^{+}$ \\

$1^{3}\Sigma^{+}$ & $a'^{3}\Sigma^{+}$ \\

$1^{5}\Sigma^{+}$ & no conventional letter label \\

$2^{1}\Sigma^{+}$, inner well & $B^{1}\Sigma^{+}$ \\

$2^{1}\Sigma^{+}$, outer well & $D'^{1}\Sigma^{+}$ \\

$2^{3}\Sigma^{+}$ & $b^{3}\Sigma^{+}$ \\

$2^{5}\Sigma^{+}$ & no conventional letter label \\

$1^{1}\Sigma^{-}$ & $I^{1}\Sigma^{-}$ \\

$1^{3}\Sigma^{-}$ & $e^{3}\Sigma^{-}$ \\

$1^{5}\Sigma^{-}$ & no conventional letter label \\

$1^{1}\Pi$ & $A^{1}\Pi$ \\

$1^{3}\Pi$ & $a^{3}\Pi$ \\

$1^{5}\Pi$ & no conventional letter label \\

$2^{1}\Pi$ & $E^{1}\Pi$ \\

$2^{3}\Pi$ & $k^{3}\Pi$ \\

$2^{5}\Pi$ & no conventional letter label \\

$1^{1}\Delta$ & $D^{1}\Delta$ \\

$1^{3}\Delta$ & $d^{3}\Delta$ \\

$1^{5}\Delta$ & no conventional letter label \\
\hline
\end{tabular}
%\end{table*}

\quad \\

\noindent The two conventional labels associated with the $2^{1}\Sigma^{+}$ adiabatic state refer to different regions of its double-well potential-energy curve: the inner well has predominantly $B^{1}\Sigma^{+}$ character, whereas the outer well has predominantly $D'^{1}\Sigma^{+}$ character. No established conventional letter labels are available for the quintet states, which are therefore identified exclusively using numerical notation.\\

\quad \\

\bibliography{CO_Jasik_Ikuesan_Bulut_Ryzner_Hakalla_Sienkiewicz}


\end{document}